\documentclass[12pt]{article}
\usepackage{mathrsfs}
\usepackage{natbib,graphicx,setspace,lscape,longtable}
\usepackage{stmaryrd}
\usepackage{natbib,epsfig,graphicx}
\usepackage{mathrsfs,amsmath,amsthm,amssymb,color}
\usepackage{enumerate}
\usepackage{multirow}
\bibpunct{(}{)}{;}{a}{,}{,}

\newtheorem{theorem}{Theorem}
\newtheorem{lemma}{Lemma}
\newtheorem{proposition}{Proposition}
\newtheorem{remark}{Remark}
 \renewcommand{\baselinestretch}{1}

\def\beq{\begin{equation}}
\def\eeq{\end{equation}}
\def\beqr{\begin{eqnarray}}
\def\eeqr{\end{eqnarray}}
\def\beqrs{\begin{eqnarray*}}
\def\eeqrs{\end{eqnarray*}}
\def\bet{\begin{theorem}}
\def\eet{\end{theorem}}
\def\bel{\begin{lemma}}
\def\eel{\end{lemma}}
\def\bep{\begin{proposition}}
\def\eep{\end{proposition}}
\def\bg{\begin{figure}[tbph]\begin{center}}
\def\eg{\end{center}\end{figure}}

\def\bc{\begin{center}}
\def\ec{\end{center}}

\def\bSigma{\mathbf{\Sigma}}

\def\blu{\mathbf{u}}
\def\blf{\mathbf{f}}
\def\blb{\mathbf{b}}
\def\bX{\mathbf{X}}

\def\bB{\mathbf{B}}
\def\mB{\mathbb{B}}

\def\var{\mathrm{var}}
\def\cov{\mathrm{cov}}

\renewcommand{\baselinestretch}{1.6}
\numberwithin{equation}{section}

\begin{document}
\title{ Estimation for ultra-high dimensional factor model: a pivotal variable detection-based approach}
\author{Junlong Zhao, Hongyu Zhao and Lixing Zhu
\footnote{The corresponding email: lzhu@hkbu.edu.hk. Junlong Zhao was supported by National Science Foundation of China, (No. 11471030, 11101022) and  Foundation of the Ministry of Education of China for Youths (No. 10YJC910013).   Lixng Zhu was supported by a  GRF grant from the University Grants Council of Hong Kong.}\\
\small\it Beihang University, LMIB of the Ministry of Education, China\\
\small\it Yale University, USA\\
\small\it Hong Kong Baptist University, Hong Kong
}
\date{}
\maketitle
\begin{abstract}
For factor model,  the involved covariance matrix often has no row sparse structure because the common factors may lead some variables to strongly associate with many others. Under the ultra-high dimensional paradigm, this feature causes existing methods for sparse covariance matrix in the literature not directly applicable.
In this paper, for general covariance matrix,  a novel approach to detect these variables that is called  the pivotal variables is suggested.
 Then,   two-stage estimation procedures are proposed to handle ultra-high dimensionality in factor model. In these procedures,  pivotal variable detection is performed as a screening step and then  existing approaches are applied to refine the working model. The estimation efficiency can be promoted under weaker assumptions on the model structure.
Simulations are conducted to examine the performance of the new method and a real dataset is analysed for illustration.
\end{abstract}

 \newpage

\baselineskip = 8.5mm
\renewcommand{\baselinestretch}{1.6}
\section{Introduction}\label{sec1}

Consider the  factor model in the form: for  $k=1,\cdots, n$
 \beq\label{nonsp_factormodel}
 \bX_k=\bB\blf_k+\blu_k,
 \eeq
 where $\bX_k=(X_{k1},\cdots, X_{kp})^T\in R^p$ are $i.i.d.$ random vectors,
$\bB\in R^{ p\times K}$ is the   loading matrix of   rank $K$ with $K$ being fixed and small, $\blf_k\in R^{K\times 1}$ is the factor vector, and $\blu_k\in R^{p\times 1}$.    For identifiability,  assume that $\cov(\blf_k, \blu_k)=0$, $\cov(\blf_k)=I_K$, $\bSigma_u=\cov(\blu_k)$ is  sparse  and $\bB^T\bB$ is the $K\times K$ diagonal matrix.  Then the covariance of $\bX_k$ has the form
$\bSigma=\bB\bB^T+\bSigma_u$. During the
 last decade, many  works   on the  inference of the  factor model has been
developed, such as, Stock and Watson (1998, 2002), Bai and Ng (2002), Bai (2003), Bai and Li (2012), Fan (2011, 2013), Luo (2011)  among others.

  When $\bB$ is nonsparse, the common factors $\blf_k$ can affect many or even all $X_{kj}, 1\le j\le  p$.  Consequently, although   $\bSigma_u$ is sparse in this model, $\bSigma$ is nonsparse in rows. See  Luo (2011) and Fan, et al (2013) for example. Thus, existing approaches in the literature may not be feasible to estimate $\bSigma$ and $\bB$. To estimate $\bSigma$, Luo (2011) suggested a LOw Rank and sparsE Covariance
 (LOREC) when $p/n \rightarrow 0$ as $n \to \infty$, and Fan, et al (2013) considered the conditional sparsity model and proposed a principal orthogonal complement thresholding method (POET) when $p/n^2 \rightarrow 0$. Interestingly, because of the special structure of the factor model, in case $\bSigma$ is sparse such that $n^{1/2}/p\rightarrow 0$ does not hold, the POET estimate cannot be consistent. Both of them cannot handle ultra-high dimension.

On the other hand, when  $p$ is very large, it is more often the case that  the  common factors $\blf_k$     affect  $s_0(p)$ components of   $\bX_k$ where $s_0(p)$ can be large, but compared with $p$,  is still relatively small. 
The matrix $\bSigma$ is dense in these rows(columns) and is sparse in the others. Therefore, when we can efficiently detect these rows, the estimation will become much easier in ultra-high dimensional scenarios.

Therefore, we suggest a novel approach to detect the variables that makes the corresponding rows dense.  The method is for general covariance matrix estimation.  It is worthwhile to mention that for covariance matrix estimation, row sparsity is commonly assumed, see  Bickel  and Levina (2008), Rothman,  Levina and Zhu (2009), Cai and Liu (2011) and  Ravikumar et al (2011). However, in some applications,  this assumption  is restrictive. Variables may have significant  differences in their behaviors.    Some variables are correlated   with many others,  while the rest are only related  to  a few.    Consequently, $\bSigma$ can be  dense in some rows and sparse in the others.  Consider  the personal relation as an example: if each person is  treated  as a  variable and  two people are related if they know each other.   Then  some people, e.g.  the  public figures,   may be related with  many others, while most of the others are related with only a few persons.
  Also  in   citation analysis with  each  article or book being viewed as a variable, some articles or books  are  cited by many others, while most   are much less cited.
In this paper, we  consider another assumption to indicate pivotal variables. That is, there exists  an index set $J\subset\{1,\cdots, p\}$ and $J^c=\{1,\cdots, p\}\backslash J$,  the rows or  columns of $\bSigma$ with  indices in  $J$ may  be nonsparse whereas those with indices in $J^c$ are sparse. The detail  is given in Section~2.
Variables corresponding to the rows that are  nonsparse are  called  the \emph{pivotal variables} whereas variables corresponding to the sparse rows are called the \emph{non-pivotal varibles}.  we investigate the estimation for the factor model~(\ref{nonsp_factormodel}) when $p$ is ultra-high. In Section~2, we give a method to detect the pivotal variables and a ridge ratio method is suggested to estimate the number of those variables.

%
%
%

In Section~3, the pivotal variable detection (PVD) to the factor model~(\ref{nonsp_factormodel}) is first performed to reduce the estimation difficulty. An algorithm to estimate the covariance matrix $\bSigma $ is proposed in a generic structure. As POET (Fan, et al 2013) and LOw Rank and sparsE Covariance
(LOREC, Luo 2011) are two promising estimation methods for the factor model with relatively high, but not ultra-high dimension $p$, we then in Sections~4 and 5 separately discuss the PVD-based POET and LOREC to show the importance of PVD for us to have more efficient estimation procedures for the factor models when $p$  can be ultra-high. Numerical studies are presented  in Section~6.

Introduce some notations first.  For  matrix $A$ of dimension $p\times p$ and index sets $I_1$ and $I_2$,  write respectively $A_{I_1I_2}$  as the sub-matrix of $A$ with rows  $I_1$ and columns $I_2$;  $A_{I_1\cdot}, A_{\cdot I_2}$ as the sub-matrices consisting of $I_1$ rows and $I_2$ columns. In particular, the sub-matrix of matrix  $\bSigma_u$  is denoted as $\bSigma_{u,I_1I_2}$, $\|A\|_1$,  $\|A\|$ and $\|A\|_F$ respectively as the $\ell_1$ norm,  operator norm,  and Frobenius norm  of $A$.  For any set $I$, $|I|$ denotes the cardinality of $I$.   For a square matrix $A$, $\lambda_{\min}(A)$ denotes the minimum eigenvalue of $A$.
In addition,  $c_i$ and $C_i$ stand for constants.

\section{Pivotal  variable detection in  high dimensional covariance matrix estimation}

\subsection{Identification of pivotal variables}
Consider the identification of pivotal variables first.
Assume the following conditions to distinguish between pivotal and non-pivotal variables. Let $J$ be the index set  of pivotal variables with cardinality  $|J|=s_0(p)$. Let $r_i=\sum\limits_{j=1}^p \sigma_{ij}^2/p$, $1\le i\le  p$ and $q_n=\sqrt{(\log p)^5/n}$.

\begin{description}
\item[] ({\bf A1})  For some constant $0<\kappa<\infty$, $\kappa^{-1}\le r_i/c_p \le \kappa$ uniformly for $i\in J$, and $ \max\limits_{i\notin J} r_i=O(\delta_p)$. Moreover,  it holds that  $q_n=O(c_p^2)$ and   $\delta_p=o(q_n)$.

\end{description}

\begin{remark}
Condition~(A1) means that $q_n^{-1}c_p^2=O(1)$ or $\infty$ and $q_n^{-1}\delta_p=o(1)$.  Since $q_n$ will be set to converge to 0,  Condition (A1)  includes the case of  $c_p=O(1)$ or $c_p\rightarrow \infty$.  It also allows $c_p\rightarrow 0$ but the rate should not be faster than   $\sqrt{q_n}$.  This condition is to distinguish between  those  $r_i$'s corresponding to  pivotal variables and nonpivotal variables through different rates.
For the    approximate factor model  where  $\bSigma=\bB\bB^T+\bSigma_u$, Fan et al (2013) assumed   that  $p^{-1}\lambda_{\min}(\bB^T\bB)>c>0$ for some constant $c$. In practice,  this assumption may fail when the common factor $\blf_k$  only affects part of variables.    In Section~3, we show that  (A1) can still hold though  $p^{-1}\lambda_{\min}(\bB^T\bB)>c>0$ fails. In this case,  the pivotal variable detection is helpful to get good estimate. Details are referred to Section~3.

\end{remark}

Recall that  $\bX_k=(X_{k1},\cdots X_{kp})^T\in R^p, k=1,\cdots,n, $ are $i.i.d.$ observations of $\bX$. Let
 $$\hat{r}_i=\sum\limits_{j=1}^p \hat\sigma_{ij}^2/p,$$
 where $\hat\sigma_{ij}=n^{-1}\sum\limits_{k=1}^n (X_{ki}-\bar{X}_i)(X_{kj}-\bar{X}_j)$ and $\bar{X}_i=n^{-1}\sum\limits_{k=1}^n X_{ki}, 1\le i, j\le p$.
 Two  conditions are assumed below:
\begin{description}
\item[(\bf{A2})]   $\log p=o(n^{1/5})$,  $n^{\epsilon_0}=o(p)$ for some constant $\epsilon_0>0$  and there exists  $T_0>0$, such that  $\sup\limits_{1\le j\le p}E\exp(X_{kj}^2/t)\le T_1<\infty$ for any $t>T_0^2$.
\item[(\bf{A3})] Let $\theta_{ij}=\var(X_{ki}X_{kj})$. $\max\limits_{1\le i,j\le p}\theta_{ij}:=\theta_0<\infty$, and
$\max\limits_{1\le i\le p}\sigma_{ii}<\sigma_0<\infty$.
\end{description}

 Condition (A2) means that $p$ has order lower than $\exp(n^{1/5})$ but higher than $n^{\epsilon_0}$ for some $\epsilon_0>0$.
 In high dimensional setting where  $p$ is usually significantly larger than $n$, $n^{\epsilon_0}=o(p)$ holds obviously  with $\epsilon_0=1$.
 $p<n$ is also allowed when  $\epsilon_0<1$. When $p$ is fixed, pivotal variable detection makes less sense, we will not discuss this scenario in this paper.  The following theorem  states the consistency of  $\hat r_i$ of $r_i$.

\bet\label{theo1}
Under Conditions~(A2) and (A3),  we have
$$P\left(\max_{1\le i\le p} |\hat r_i-r_i|>C_0q_n\right)=O(p^{-\delta_0})$$
where $M>1+\epsilon_0^{-1}+\delta_0$ with $\delta_0$ being sufficiently small and $C_0$ is a constant depending  on $M$ and $T_0$.
\eet

\begin{remark}
From the proof in the supplement, we see that  $C_0>24M^2T_2^2$, where $T_2=\max\limits_{1\le i\ne j\le p}\|X_{ki}X_{kj}\|_{\psi_1}$ being a constant depending on $T_0$ and $\|\cdot\|_{\psi_1}$ is the $\psi_1$ norm \citep{Vershynin:2011}.  It can be shown that  $T_2\le 2T_0^2$. Note that  the value of $C_0$ here is only an upper bound. Since $T_0$ is generally unknown, $C_0$ is also an unknown constant. Thus, this result is mainly for theoretical justification. However, in Subsection~2.2 below for estimating the number of pivotal variables by a ridge ratio method, we can recommend a value of ridge for practical use without involving this unknown $C_0$.
\end{remark}
  Combining this result with Theorem 1 and Condition (A1), we can  shows that   the maximum of   $\hat r_i$  with $i\in J^c$  is significantly less than minimum of $\hat r_i$   with $i\in J$. This   provides a foundation  for the identification of  pivotal variables.
\bet\label{prop1}
  Under Conditions~(A1)-(A3) stated above, we have  $\max\limits_{i\in J^c} \hat r_i/\min\limits_{i\in J} \hat r_i=o_p(1)$.
\eet
We  can see from this proposition that, as $n$ being large, the indices with  larger values of $\hat r_i$ are  associated with  pivotal variables and those with smaller values of $\hat r_i$ are associated with non-pivotal variables.  Sort  $\hat r_i, 1\le i\le p$ in decreasing order, denoted  as $\hat r_{(1)}\ge \hat r_{(2)}\ge \cdots \ge \hat r_{(p)}$.   Then  the indices associated with  $\hat r_{(1)},\cdots, \hat r_{(s_0(p))}$ can be  the estimate of $J$, where  $s_0(p)=|J|$.  However, $s_0(p)$ is unknown. In the following subsection, we will develop  an effective  method to estimate $s_0(p)$.

\subsection{Consistent  estimate of the number of pivotal variables}

 In this section we consider  estimating $s_0(p)$. A ratio estimate that is based on $\hat r_i$'s is suggested.  It can be used a criterion to estimate $s_0(p)$ because of the following observation.   Without loss of generality,  assume that $J=\{1,\cdots, s_0(p)\}$ and   that the values of $r_i, i\in J$  have the  decreasing  order  $r_1\ge  r_2\ge \cdots \ge r_{s_0(p)}$. At the population level, $1\ge r_{i+1}/r_i>C>0$ for a positive constant $C$ when $1\le i<s_0(p)$ and when $i=s_0(p)$, $r_{i+1}/r_i \approx 0$.
 In other words, at the value of $i=s_0(p)$, the ratio has a clear dropdown in value. Although when $i>s_0(p)$,  some ratios  may be close to  $0/0$, we can add a ridge to make all the ratios well defined. That is, $(r_{i+1}+l)/(r_i+l)$ for a very small positive value $l$.  Thus, we have, for $i<s_0(p)$ and $j>s_0(p)$, as long as $l$ is small enough (at the sample level, we let it go to zero at certain rate later),
 $$
 (r_{i+1}+l)/(r_i+l)>l/(r_{s_0(p)}+l)<1=l/l \approx (r_{j+1}+l)/(r_j+l).
 $$
This means that $s_0(p)$ is the minimizer of the ratios over all $i$ with $1\le i\le p$. At the sample level, we can replace $r_i$ by the corresponding estimates.
   Recall that $\hat r_{(1)}\ge \hat r_{(2)}\ge \cdots\ge \hat r_{(p)}$ is the decreasing order of  $\hat r_i, i=1,\cdots, p$. The sample criterion is
$$ R_i=\frac{\hat r_{(i+1)}+l_n}{\hat r_{(i)} +l_n},\ \ \ \ \    i=1,\cdots, p-1,$$
where  $l_n\rightarrow 0$ to be specified below. The principle of choosing $l_n$ is as follows. First, $l_n$ goes to zero such that the minimum of $R_i$ can go to zero, and second, the convergence rate of $l_n$ to zero should be slower than $r_{s_0(p)+1}$ to zero such that $l_n$ can be a dominating factor such that $R_i$ for $i>s_0(p)$ converge to $1$.  Then  $s_0(p)$ and $J$ can respectively be  estimated  by
\begin{eqnarray}\label{eq2.1}
\hat s_0(p)=\arg\min\limits_{1\le i\le p} R_i \quad \mbox{and}\quad
\hat J=\{i: \hat r_i\ge \hat r_{(\hat s_0(p))}\}
\end{eqnarray}
This criterion is in spirit similar to that in Xia, Xu and Zhu (2014). The consistency of $\hat s_0(p)$ and $\hat J$ is stated in the following.
\bet\label{theo3}
Under Conditions~(A1)-(A3) in Subsection 2.1,  as  $l_n=O([(\log p)^{5}/n]^{\delta_1})$, with $\delta_1\in (\frac{1}{4},\frac{1}{2})$,   we have
$P(\hat s_0(p)=s_0(p))\rightarrow 1$ and $P(\hat J=J)\rightarrow 1.$
\eet

Theorem \ref{theo3} imposes  a constraint on the order of $l_n$.  A simple choice can be  $l_n=[(\log p)^5/n]^{3/8}$, which is used in our simulations in Section~6.

\section{  Application to  factor model}

\subsection{Factor model}
Recall the factor model~(\ref{nonsp_factormodel}):
 \beq
 \bX_k=\bB\blf_k+\blu_k,\nonumber
 \eeq
   where $\blu_k\in R^{p\times 1},  \bX_k\in R^{p\times 1}$, $\blf_k\in R^{K\times 1}$ and $\bB$ is a matrix of dimension $p\times K$ and  $K$ is an unknown small integer. In addition,  assume that   rank$(\bB)=K$, $\cov(\blf_k)=I_K$, $\cov(\blf_k, \blu_k)=0$, $\bB^T\bB$  is a diagonal matrix and $\bSigma_u=\cov(\blu_k)$ is a sparse matrix.

Let $\mB=\bB\bB^T$.  It is easy to see that the covariance matrix of $\bX_k$ for this model has the form:
   \beq\label{cov_fac_model}
   \bSigma=\bB\bB^T+\bSigma_u=\mB+\bSigma_u.
   \eeq
In Fan, et al (2013) and Luo (2011), the rows of the loading matrix $\bB$ are nonzero.     Thus, the common factors $\blf_k$ could have impact for many or even all the variables $X_{kj},1\le j\le p$.  We  call (\ref{nonsp_factormodel})  the \emph{ nonsparse factor model}.
 A natural way to estimate the loading matrix $\bB$ and the factors is through estimating $\bSigma$.  However, it is not easy unless the  dimension $p$ is not ultra-high. As we pointed out in the introduction, Luo (2011) requires $p/n\rightarrow 0$  and Fan et al (2013) requires $p/n^2\rightarrow 0$ and $n^{1/2}/p\rightarrow 0$.

On the other hand, in factor analysis, it is often the case that   many rows of the loading matrix $\bB$ have very small  or zero  values. In other words, the factors can  have impact for part of variables and thus although $\bB$ is not sparse, the  number of variable affected by the factors is not very large compared with the  ultra-high dimension $p$. Therefore, a direct way to reduce dimensionality is to first identify those variables who are affected by the factors associated with $\bB$. This way offers us a separation between two types of variables who respectively are affected  and are not affected by the factors. We  apply the pivotal variable detection for this purpose. When the number of pivotal variables $s_0(p)$ is much smaller than the original dimension $p$, we can then use either the method in Fan et al (2013) or that in Luo (2011) to estimate $\blf_k, \bB$ and $\bSigma_u$ in a dimension-reducing model.

   Assume  that there exists a subset $J\subseteq\{1,\cdots, p\}$ such that  the rows of $\bB$ with the index set $J^c=\{1,\cdots, p\}\setminus J$ are 0. That is,  letting $\bB=(\blb_1,\cdots\blb_p)^T=(b_{ij})$, then $\blb_j=0$ for $j\in J^c$. Write $\bB_{J \bullet}$ as the matrix consisting of the rows with the index set $J$ and  $\bB_{J^c \bullet}$ as the matrix with the rows associated with the index set $J^c$.
By the definition of  $\mB$,  the factor model can be rewritten as
 \beqr\label{sp_fact_model}
  \bX_{kJ}=\bB_{J\bullet}\blf_{k}+\blu_{kJ},\ \ \ \    \bX_{kJ^c}=\blu_{kJ^c}
  \eeqr
 where $\bX_{kJ}$ is the sub-vector of $\bX_k$ with the index $J$ and $ \blu_{kJ}$ is defined similarly. Since the factor loading $\bB$ is sparse, this model is  called the \emph{sparse factor model}. For  model (\ref{sp_fact_model}),  it is easy to see that
 $\bSigma=\mB+\bSigma$ in which the submatrices $\mB_{J^cJ}, \mB_{JJ^c}, \mB_{J^cJ^c}$ of $\mB$ are zero matrices.

 To estimate  the corresponding $\blf_k$, $\bB_{J, \bullet}$ and $\bSigma_{u, J J}$ that is the submatrix of $\bSigma_{u}$ with the index set $J$, we first  identify the index set $J$. After that, sophisticated methods in the literature can be applied.  For the matrix $\bSigma_{u, J^c J^c}$ associated with $\blu_{kJ^c}$, we can estimate it by existing methods. we will discuss it in detail later.

 { To accommodate the methodology development in this section, we first state the conditions and results in Fan et al (2013) for   principal orthogonal complement thresholding (POET).
Denote $\lambda_{p,\bB}=p^{-1}\lambda_{\min}(\bB^T\bB)$.
The key condition for POET to work is  the pervasive assumption (Assumption~1 in Fan et al (2013)):
\beq\label{Pervasive}
\lambda_{p,\bB}>c>0\ \ \ \mbox{and} \ \ \  p^{-1}\|\bSigma_u\|\rightarrow 0.
\eeq
Under this condition, $\bSigma$ is a spike matrix, of which the first $K$ largest eigenvalues of $\bSigma$ increases to infinity at the rate of  order $O(p)$. This condition leads the principal component analysis (PCA) to work on constructing  a consistent  estimate of $\mathrm{span}(\bB)$.  If  this  condition fails, the POET estimate may be inconsistent.}

  However,   for the factor model~(\ref{sp_fact_model})  the pervasive assumption  (\ref{Pervasive}) may fail to hold.
 Recall that  $\bB=(b_{ij})$.  Let $b_{\max}=\max\limits_{i,j} |b_{ij}|$ and suppose that  $b_{\max}<\infty$.  As $|J|/p\rightarrow 0$,   then $\lambda_{p,\bB}\rightarrow 0$ and (\ref{Pervasive}) fails. As a result, POET may not guarantee the consistency of the estimates of  $\mathrm{span}(\bB)$, $\bSigma_u$ and $\bSigma$.
As pointed out by Fan et al (2013), the more variables  the common factors can  affect, the stronger  their  signals  are and easier they can be detected. In other words, in the case of $|J|$ being small, such as $|J|/p\rightarrow 0$, the signals of the common factors are  relatively weak and the  detection for them becomes relatively difficult.

  Note that the  rows and columns of $\bSigma$ with index $J$ are less sparse in  model~(\ref{sp_fact_model}). Then our idea is first to estimate  the index $J$ by the pivotal variable detection method. Afterwards,   we can estimate  $\bSigma$ by separately treating $\bX_{kJ}$ and $\bX_{kJ^c}$. Details are presented in Section~3.2.
To detect $J$ correctly,  Condition~(A1) in Section~2.1 is required.  For  model (\ref{sp_fact_model}),  it is easy to see that
$r_i=p^{-1} \|\bB_{i.}\bB^T+\bSigma_{u,i.}\|^2$ for  $i\in J$ and  $r_i=p^{-1}\|\bSigma_{u,i.}\|^2, i\in J^c$.   Now, we give  sufficient conditions for Condition~(A1)    by  imposing an assumption on $\bB$, such that $\|\bB_{i.}\bB\|^2$  dominates  $\|\bSigma_{u,i.}\|^2$. Clearly this  condition is not the weakest but is   easy to understand.
\bep\label{Prop3} For  model~(\ref{sp_fact_model}), suppose that $q_n=O(c_p^2)$ and $\delta_p=o(q_n)$, and
\begin{itemize}
 \item[(1)] $\frac{\lambda_{\max}(\bB^T\bB)}{\lambda_{\min}(\bB^T\bB)}=O(1)$,   $ \sqrt{p^{-1}c_p}/\lambda_{p,\bB}=O(1)$,
  \item[(2)]  $\|\bSigma_u\|=o(\sqrt{p\delta_p})$ or  $\max_{i\in J^c}\|\bSigma_{u,i.}\|=o(\sqrt{p\delta_p})$.
 \end{itemize}
   Then Condition (A1) in Subsection~2.1 holds.
\eep

Here the assumption $\frac{\lambda_{\max}(\bB^T\bB)}{\lambda_{\min}(\bB^T\bB)}=O(1)$ is used to   guarantee  that all $r_i$ with $i\in J$ have the same magnitude.  Recall that  $b_{\max}=\max\limits_{i,j} |b_{ij}|$.  As $b_{\max} <\infty$,  we can show that  $\lambda_{p,\bB}>c>0$ in Fan et al (2013) implies $\frac{\lambda_{\max}(\bB^T\bB)}{\lambda_{\min}(\bB^T\bB)}=O(1)$.   Proposition~\ref{Prop3} relaxes their assumption such that      $\lambda_{p,\bB}$ can be $O(1)$ or even tends to 0 at a rate slower than $\sqrt{c_p/p}$.      In addition, note that for any $0<q\le 1$,  $\|\bSigma_{u,i.}\|\le \|\bSigma_{u,i.}\|_q$.  If $\bSigma_u$ satisfies the row  sparsity  with $\max_{1\le i\le p}\|\bSigma_{u,i.}\|_q=o(\sqrt{p\delta_p})$,  condition (2) here  holds naturally.

 Recall that $\bSigma_u=(\sigma_{u,ij})$,  $\bB=(b_{ij})$ and $\blu_k=(u_{k1},\cdots, u_{kp})^T\in R^p, 1\le k\le n$.
   We give some  conditions below such that $J$ can be consistently estimated.
\bet\label{theo6} Suppose that  (i) Condition  (A1) in Subsection~2.1 holds, $\log p=o(n^{1/5})$ and $\blf_k, u_{kj}, 1\le j\le p $ are  subgaussian variables;  (ii) for some constant  $C>0$ such that  $b_{\max}$, $\max\limits_{1\le i,j\le p}|\sigma_{u,ij}|$, $\|\blf_k\|_{\psi_2}$ and $\max\limits_{1\le j\le p}\|u_{kj}\|_{\psi_2}$ are bounded above by $C$.
Then we have
$$P(\hat J=J)\rightarrow 1,$$
where $\hat J$ is the estimate of $J$ obtained by the pivotal detection method in Section~2 and the definition of $\|\cdot\|_{\psi_2}$ is referred to \cite{Vershynin:2011}.
\eet

\subsection{Covariance matrix estimation}
 We are now in the position to investigate the covariance matrix estimation for   model (\ref{sp_fact_model}). Recall that   the covariance matrix has the form  $\bSigma=\mB+\bSigma_u$ with $\mB_{J^cJ}, \mB_{JJ^c}, \mB_{J^cJ^c}$ being  zero matrices. The blocks of its covariance matrix have the following  specific  structures:
 \beqr\label{(3.1)}
 \bSigma_{JJ^c}&=&\cov(\bX_{iJ}, \bX_{iJ^c})=\cov(\blu_{iJ}, \blu_{iJ^c})=\bSigma_{u,JJ^c},\nonumber\\
    \bSigma_{J^cJ^c}&=&\cov(\bX_{iJ^c})=\cov(\blu_{iJ^c})=\bSigma_{u,J^cJ^c},\\
 \bSigma_{JJ}&=&\bB_{J\cdot}\bB_{J\cdot}^T+\bSigma_{u,JJ}=\mB_{JJ}+\bSigma_{u,JJ},\nonumber
 \eeqr
 where $\bSigma_{u,JJ^c}$ is the submatrix of $\bSigma_u$ with indices of row $J$ and  column $J^c$; other quantities are defined similarly. %
Note that $\bSigma_{u}$ and $\bSigma$  have the same block matrices  with indexes  $(J,J^c), (J^c,J)$ and $(J^c,J^c)$ respectively.  Since $\bSigma_u$ is sparse, the three block matrices $\bSigma_{JJ^c}, \bSigma_{J^cJ}, \bSigma_{J^cJ^c}$ of  $\bSigma$ are also sparse.  Note that the pivotal variable detection can be applied to identify and consistently estimate the index set $J$, we can then have an estimation strategy to separately estimate these four block matrices that are associated with the index sets $(J,J)$, $(J, J^c)$, $(J^c, J)$, and $(J^c, J^c)$. First, we can apply the existing thresholding penalty method (e.g. Rothman,  Levina and Zhu (2009)) on the corresponding block matrices $\hat\bSigma_{\hat J\hat J^c}$, $\hat\bSigma_{\hat J^c \hat J}$ and $\hat\bSigma_{\hat J^c \hat J^c}$ of the sample covariance matrix  $\hat\bSigma=n^{-1}\sum_{k=1}^n(\bX_{k}-\bar \bX)(\bX_k-\bar \bX)^T$ where $\bar \bX=n^{-1}\sum_{k=1}^n \bX_k$.

Second, we consider how to estimate  the block matrix $\bSigma_{JJ}$ which is the sum of a low rank matrix $\mB_{JJ}$ and the sparse matrix $\bSigma_{u,JJ}$.  Note that $\bSigma_{JJ}$ is the covariance matrix of the submodel  $\bX_{k J}=\bB_{J\cdot}\blf_k+\blu_{kJ}$, which is a nonsparse factor model.  Therefore, existing methods developed for nonsparse factor model can be used to estimate  $\bSigma_{JJ}$ by the data $\bX_{k\hat J}, 1\le k\le n$. Since the dimension of $\bX_{k \hat J}$ is $s_0(p)$ much smaller than $p$, estimating this sub-model becomes a problem with small or moderate dimension.

The estimation  procedure is then summarised to the following four   steps.
\begin{description}
\item[]\emph{Step 1}.  Apply  the pivotal variable detection method in Section~2 to consistently estimate  the index set $J$. The estimate is defined as $\hat J$;
\item[] \emph{Step 2}.  Apply an existing method to obtain estimates that are based on the data $\bX_{k\hat J}, k=1,\cdots, n$. In the following two sections, we will give the details about principal orthogonal complement thresholding ( POET, Fan, et al 2013), and low rank and sparse covariance
(LOREC, Luo 2011), and the comparisons with these two methods when our method is combined with them.
\item[] \emph{Step 3}.  Together with  the results in Step 2, use the  thresholding method to define an estimate $\hat\bSigma_u^\tau$ of $\bSigma_u$, see 
    Rothman,  Levina and Zhu (2009) and Cai and Liu (2011).
\item[] \emph{Step 4}.   $\bSigma$ is   estimated by $\hat\bSigma^\tau=\hat{\mathbb{B}}+\hat\bSigma_u^\tau $, where  $\hat{\mathbb{B}}_{\hat J^c \hat J}=0$, $\hat{\mathbb{B}}_{\hat J \hat J^c}=0$ and $\hat{\mathbb{B}}_{\hat J^c \hat J^c}=0$.
\end{description}

Now we give some  discussions on Step 2.
Many methods have been developed to estimate the covariance matrix in the model $\bX_k=\bB \blf_k+\blu_k$ without the sparse assumption $\bB_{J^c\cdot}=0$.   As was pointed out before, estimating this   model  requires   strong  assumptions, especially on $p$, e.g.   $p/n\rightarrow 0$ in Luo (2011).  However,  our method avoids this difficulty because in Step~2,  we  consider the factor model (\ref{sp_fact_model}) rather than the full model~(\ref{nonsp_factormodel}), which only involves $s_0(p)$ covariates rather than the original $p$ covariates. When $s_0(p)$ is  small, and then estimation can be much easier and efficient.

 In principle, many existing methods can be applied in Step 2. But to make estimation easier and more efficient, the method we  use for this purpose highly depends on specific structure of covariance matrix. There are several proposals in the literature such as  Chandrasekaran et al. (2010), Agarwal et al. (2011), Fan et al. (2013) and  Luo (2011).
 In this paper, we adopt two methods in Step~2:  POET (Fan et al, 2013) and LOREC (Luo, 2011).  The  pivotal variable detection based POET and LOREC  are respectively denoted as PVD-based POET and PVD-based OREC.   In Sections~4 and 5,  we respectively compare PVD-based POET and POET; and PVD-based OREC  and LOREC.  Theoretical results in Sections~4 and 5 and numerical results in Section~6 show that  our method can improve the performances of POET and LOREC  significantly when $s_0(p)$ is relatively small compared with $p$.

\section{PVD-based LOREC}

\subsection{A brief review of LOREC}

LOw Rank and sparsE Covariance
estimator (LOREC, Luo, 2011) deals with the following covariance matrix $\Sigma^*$ with the form
$\Sigma^*=L^*+S^*,$
 where $L^*$ is a low rank matrix and $S^*$ is a sparse matrix. This includes the  factor model~(\ref{cov_fac_model}) as a special case with $\Sigma^*=\bSigma$, $L^*=\bB\bB^T=\mB$ and $S^*=\bSigma_{u}$.    To get an estimate
LOREC solves the following optimization problem:
\beq\label{LOREC}
\min_{L,S}  \frac{1}{2} \|L + S -\hat\bSigma\|_F^2 + \lambda \|L\|_* + \rho\|S\|_1
\eeq
where $\hat\bSigma$ is the sample covariance matrix, $\|A\|_*$ is the nuclear (trace) norm of  matrix $A$, $\lambda$ and $\rho$ are tuning parameters.
Let $\hat\bSigma_L$ denote the LOREC estimate of $\bSigma$. This estimation procedure is general, and does not take care of the sparsity of $\mB$.

We first give some notations that were introduced in Luo (2011). For any matrix  $M\in R^{p\times p}$ with the SVD decomposition $M=UDV^T$ with $U\in R^{p\times r}, V\in R^{p\times r}$, and a diagonal matrix $D\in R^{r\times r}$. Define the tangent spaces
$$\Omega(M)=\{N\in R^{p\times p}| \mbox{support}(N)\subseteq  \mbox{support}(M)\},$$
  $$T(M)=\{UY_1^T+Y_2V^T|Y_1,Y_2\in R^{p\times r}\}.$$
 Define respectively the coherence measures of $\Omega(M)$ and $T(M)$ by
 $$\xi(T(M))=\max\limits_{\substack N\in T(M), \|N\|_2\le 1} \|N\|_\infty, \ \ \ \ \mu(\Omega(M))=\max\limits_{N\in \Omega(M),\|N\|_\infty\le 1}\|N\|_2.$$
Typically
a matrix $M$ with incoherent row/column spaces would have $\xi(T(M)) \le 1$, and $\xi(T(M)) = 1$ if the row/column spaces of M contain a standard basis
vector. Note that $\xi(T(M))$
can be as small as $O(\sqrt{r/p})$ for a rank-$r$ matrix $M \in  R^{p\times p}$.   Detailed discussions of the above quantities
$\Omega(M), T(M), \xi(T(M))$ and $\mu((M))$ and their
implications can be found in Chandrasekaran et al. (2012) and  Luo (2011).
Let $\mathcal{U}(\epsilon_0)=\{M\in R^{p\times p}: 0<\epsilon_0<\lambda_{i}(M)<\epsilon_0^{-1}<\infty \},$ where $\lambda_1\ge \lambda_2\ge\cdots\ge \lambda_p$ are the singular value of $M$.  

Under certain regularity conditions,
Corollary 2 of Luo (2011) shows that
$\|\hat\bSigma_{L}-\bSigma\| =O_p(\tilde v_{1n})$,
 where
\beq\label{LOrate}\tilde v_{1n}=[\tilde s  \xi(T(\mB))+1] \max\left\{\frac{1}{\xi(T(\mB))}\sqrt{\frac{\log p}{n}}, \sqrt{\frac{p}{n}} \right\}\eeq
with  $\tilde s=\max\limits_{1\le i\le p}\sum\limits_{1\le j\le p} I(\sigma_{u,ij}\ne 0)$, and $\xi(T(\mB))$ can be bounded by $1$ and in some cases, it can be as small as $O(\sqrt{r/p})$. The details can be found in Chandrasekaran et al. (2012).
Therefore, $p/n\rightarrow 0$ is a necessary condition to guarantee the consistency of $\hat\bSigma_{L}$. In other words, LOREC can not generate a consistent estimator when $p$ is much larger than $n$ even when  $\mB$ is sparse in the model~(\ref{cov_fac_model}).

\subsection{PVD-based LOREC for   model~(\ref{sp_fact_model})}


In contrast, for the sparse factor model, the pivotal variable detection in Step~1 is  to reduce it to model~(\ref{sp_fact_model}) to make estimating easier. Let $\hat J$ be the estimate obtained by the pivotal variable detection. Then we use LOREC to estimate $\bSigma_{JJ}=\bB_J\bB_J^T+\bSigma_{u,JJ}=\mB_{JJ}+\bSigma_{u,JJ}$.
Let $\hat\bSigma_{\hat J\hat J}=n^{-1}\sum\limits_{k=1}^n(\bX_{k\hat{J}}-\bar{\bX}_{\hat J})(\bX_{k\hat{J}}-\bar{\bX}_{\hat J})^T$.
Replacing $\hat\bSigma$ with $\bSigma_{\hat J\hat J}$ in (\ref{LOREC}), we respectively define the estimates $\hat{\mathbb{B}}_{\hat J\hat J}$ and $\hat\bSigma_{u,\hat J\hat J}$ of $\mB_{JJ}$ and  $\bSigma_{u,JJ}$.

To define the final estimate of $\bSigma$, Step~4 tells us that what we need to do is to estimate the other elements in $\bSigma_{u}$. Combining the above estimate of $\bSigma_{u,JJ}$, we only need to estimate $\sigma_{u,ij}$ for either $i\notin  J$ or $j\notin  J$.    Note that  $\sigma_{u,ij}=\sigma_{ij}$ for either $i\notin J$ or $j\notin J$ by (\ref{(3.1)}). Thus, we can  use  $\hat\sigma_{ij}=n^{-1}\sum\limits_{k=1}^n (X_{ki}-\bar{X}_i)(X_{kj}-\bar{X}_j)$   an estimate of $\sigma_{ij}$ for either $i\notin \hat J$ or $j\notin \hat J$.
{
Since LOREC  estimates $\bSigma_u$ by $L_1$ penalty function,  to make a fair  comparison between the PVD-based LOREC and LOREC,  we  use the same method to estimate   $\sigma_{u,ij}$ with $i\notin \hat J\ \mbox{or}\ j\notin \hat J$.}
As a result,  the  soft  thresholding penalty function (Rothman, Levina and Zhu, 2009)  is applied to  $\hat\sigma_{ij}$ to define a sparse estimate of $\sigma_{ij}$. Let $\tau_{ij}=C\sqrt{(\log p)/n} $ for either $i\notin \hat J$ or $j\notin \hat J$. We then obtain an estimate $\hat\bSigma_{u}^\tau$ that is related to the thresholding value $\tau_{ij}$. Together with Step~4, we obtain an estimate $\hat\bSigma^\tau$ of $\bSigma$.

%

To investigate the theoretical property of the estimate,  the  following condition is  similar as that  in
Theorem~1 of Luo (2011).
\begin{description}
\item[({\bf A4})]  Let $\Omega_J=\Omega(\bSigma_{u,JJ})$ and $T_J=T(\mB_{JJ})$.  Assume that $\bSigma_{JJ}\in \mathcal{U}(\epsilon_0)$,   $\mu(\Omega_J)\xi(T_J)<1/54$, $s_0(p)<n$ and that $\lambda_n=\max(\xi(T_J)^{-1} \sqrt{(\log s_0(p))/n}, \sqrt{s_0(p)/n})$ and $\rho_n=\kappa \lambda_n$, where $\kappa\in [9\xi(T_J), 1/(6\mu(\Omega_J))]$.
\end{description}

Let $v_{1n}=s\xi(T_J)\max\left\{\frac{1}{\xi(T_J)}\sqrt{\frac{\log s_{0}(p)}{n}} , \sqrt{\frac{s_0(p)}{n}}\right\}$,  $v_{2n}=m_p \left(n^{-1}\log p\right)^{(1-q)/2}$, where  $m_p=\max\limits_{1\le i\le p}\sum\limits_{1\le j\le p}|\sigma_{u,ij}|^q$ and $0\le q< 1$, and
$s=\max_{i\in J} \sum_{ j\in J} I(\sigma_{u,ij}\ne 0)$.
 Then we have the following conclusion.

\bet\label{theo5}
 Under Conditions~(A1)--(A3) in Subsection 2.1 and (A4) stated above, for the model~(\ref{sp_fact_model})
$$\|\hat\bSigma_{u}^{\tau}-\bSigma_{u}\|=O_p(v_{1n}+v_{2n}),$$
$$\|\hat\bSigma^{\tau}-\bSigma\|=O_p(v_{1n}+v_{2n}+\lambda_n).$$
\eet

\subsection{A comparison between PVD-based LOREC  and LOREC}

We now briefly make a comparison with LOREC described in Section~4.1. The comparison consists of two parts. The first part is about the practical implementation. We notice that LOREC is a computationally intensive algorithm. In the simulations in Section~6, we will see this. In the sparse factor model, using PVD to make an initial screening is very helpful in the computational aspect. The second part is about its theoretical properties. As was stated before, the  LOREC estimate $\hat\bSigma_{L}$ of $\bSigma$ has the convergence rate $O_p(\tilde v_{1n})$, whereas our estimate has the rate of order  $v_{1n}+v_{2n}+\lambda_n$.
Note that
$$v_{1n}+\lambda_n=[s\xi(T_J)+1]\max\left\{\frac{1}{\xi(T_J)}\sqrt{\frac{\log s_{0}(p)}{n}}, \sqrt{\frac{s_0(p)}{n}}\right\}. $$
As $\bB_{J\cdot}=0$, by the definition of $\xi(T(\cdot))$, it is easy to see that  $\xi(T_J)=\xi(T(\mB))\le 1$ and has a lower bound $O(\sqrt{K/s_{0}(p)})$ (Chandrasekaran, et al, 2012).  It is obvious that  $s\le \tilde s$.
Therefore, it retains that $\lambda_{n}+v_{1n}\le \tilde v_{1n}$.
When  $m_p$ is small,  $v_{2n}$ can  also  be  dominated by  $v_{1n}+\lambda_n$  from the discussion below. These observations suggest that the PVD-based LOREC can generate an estimate with a convergence rate faster than or  equal to that of the  LOREC estimate.

Further, as was discussed, as $p/n\nrightarrow 0$, the LOREC estimate may be inconsistent.   In contrast, when  $s_0(p)$ is small, the consistency of  the PVD-based LOREC estimate can be ensured. This can be observed below.
Note that $\xi(T)\le 1$. We have  $v_{1n}\le s\sqrt{\frac{s_0(p)}{n}}=: w_{1n}$ and
\beqr
\lambda_n& \le  & O\left(\max\left\{\sqrt{\frac{s_0(p)\log s_0(p)}{nK}}, \sqrt{\frac{s_0(p)}{n}}\right\}\right)=O\left(\sqrt{\frac{s_0(p)\log s_0(p)}{n}}\right)=: w_{2n},\nonumber
 \eeqr
where  we have used the fact that $\xi(T(\mB))$ has a lower bound $c\sqrt{K/s_{0}(p)}$ for some positive constant $c$.  Therefore, as long as  $s_0(p)$,   $s$ and $m_p$ are small such that $\max\{w_{1n}, w_{2n}, v_{2n}\}\rightarrow 0$, where $v_{2n}=m_p \left(n^{-1}\log p\right)^{(1-q)/2}$,  we have $\|\hat\bSigma^{\tau}-\bSigma\|\rightarrow_p 0$. For example,  if  $\max(s, m_p)<\infty$,  and both  $(\log p)/n\rightarrow0$ and $\sqrt{[s_0(p)\log s_0(p)]/n}\rightarrow 0$,  the PVD-based LOREC estimate is consistent.

\section{PVD-based POET}
\subsection{A  brief review of POET}
 For  nonsparse factor model $\bX_k=\bB\blf_k+\blu_k, 1\le k\le n$,  the covariance matrix has the form
$$\bSigma=\bB\bB+\Sigma_u=\mB+\bSigma_u.$$
 Under Assumption (\ref{Pervasive}), $\bSigma$ is a spike matrix with the first $K$ eigenvalue significantly larger than the others.
The eigenvalue decomposition of $\hat\bSigma$ is $\hat\bSigma=\sum\limits_{i=1}^p\hat\lambda_i\hat\eta_i\hat\eta_i^T$, where $\hat\lambda_1\ge \hat\lambda_2\cdots \ge \hat\lambda_p$ are the eigenvalues and $\hat\eta_i$ are the corresponding eigenvectors.   Fan, et al (2013) showed that the  estimate $\hat K$ of $K$ can be consistent,  and  span$(\bB)$ can be consistently estimated by $\mathrm{span}(\hat\eta_1,\cdots,\hat\eta_{\hat{K}})$. Moreover, $\bB\blf_k$ and consequently   $\blu_k, k=1,\cdots, n$ can also  be consistently estimated. As a result, $\hat\bSigma_u^{\mathcal{T}}$ obtained by the thresholding method is an estimate of $\bSigma_u$. Then $\bSigma$ can be consistently estimated by $$\hat\bSigma^{\mathcal{T}}=\sum\limits_{i=1}^{\hat K}\hat\lambda_i\hat\eta_i\hat\eta_i^T+\hat\bSigma_u^{\mathcal{T}}.$$
In the above procedure, the consistency of  $\mathrm{span}(\hat\eta_1,\cdots,\hat\eta_{\hat{K}})$ is the  prerequisite for  the final estimate to be consistent. Without Assumption (\ref{Pervasive}), the consistency of $\mathrm{span}(\hat\eta_1,\cdots,\hat\eta_{\hat{K}})$ and then of the final estimate $\hat\bSigma^{\mathcal{T}}$  may be questionable. However, as was discussed in Section 3, Assumption (\ref{Pervasive}) may fail in model~(\ref{sp_fact_model}).

\subsection{PVD-based POET for  model~(\ref{sp_fact_model})}
Again, $J$ is  estimated by  $\hat J$ that is obtained in Step~1. In Step~2, POET is applied to the data $\bX_{k\hat J}, k=1,\cdots, n$  to  get an estimate   $\hat{\blu}_{k\hat J}$ of $\blu_{kJ}$, for  $k=1,\cdots, n$, and  $\hat \bB_{\hat J\bullet}$, whose column space is  an estimate of span$(\bB_{J\bullet})$ where $\bB_{J\bullet}$ means the $J \times K$ matrix consisting of the corresponding rows of $\bB$ to the index set $J$, and $\hat \bB_{\hat J\bullet}$ is defined similarly. In other words, POET is applied to  model (\ref{sp_fact_model}).

Further,
 recall that  $\hat\sigma_{ij}=n^{-1}\sum\limits_{k=1}^n (X_{ki}-\bar{X}_i)(X_{kj}-\bar{X}_j)$ and $\theta_{ij}=\var(X_{ki}X_{kj})$, $1\le i,j\le p$.  Define an estimate of $\theta_{ij}$ as
$$\hat\theta_{ij}=n^{-1}\sum\limits_{k=1}^n [X_{ki}X_{kj}-\bar X_i\bar X_j-\hat\sigma_{ij}]^2, \ \  1\le i,j \le p. $$
Let
$\omega_n=\sqrt{1/s_0(p)}+\sqrt{\log p/n}$ and define the thresholding values $\tau_{ij}=C\omega_n\sqrt{\hat\theta_{ij}}$ for $1\le i,j\le p$.
Then Steps 3 and 4 of the algorithm  can be reformulated as follows.
\begin{description}
\item[] \emph{Step 3'}.  Define the vectors  $\tilde\blu_k, k=1,\cdots, n$, such that  $\tilde\blu_{k\hat J}=\hat{\blu}_{k\hat J}$ and  $\tilde\blu_{k\hat J^c}=\bX_{k\hat J^c}$.  Apply the adaptive thresholding estimate  to data  $\tilde\blu_k, k=1,\cdots, n$ to obtain the estimate $\hat\bSigma_u^\tau$ of $\bSigma_u$, using the thresholding value $\tau_{ij}$. The reader can refer to Fan, et al (2013) for details.

\item[] \emph{Step 4'}. $\bSigma$ is   estimated by $\hat\bSigma^\tau=\hat \bB\hat \bB^T+\hat\bSigma_u^\tau$, where  $\hat\bB_{\hat J^c \bullet}=0$.
\end{description}

{
Since POET  uses the adaptive thresholding method suggested by  Cai and Liu (2011) to estimate $\bSigma_u$,  in Step 3', we also use this method such that  POET and PVD-based POET can be compared fairly. }  Again, as POET is used, we assume the following condition in which Part (iia-c) are the adapted versions of Assumptions 2 and 4 in Fan, et al (2013) in our setting.

\begin{description}
\item[(A5)]\ Assume that
\begin{description}
\item[]  (i)  $s_0(p)^{-1}\lambda_{\min}(\bB_{J\bullet}^T\bB_{J\bullet})$ is bounded away from both 0 and $\infty$ as $p\rightarrow \infty$.
\item[] (iia) There are constants $c_1$ and $c_2>0$ such that $\lambda_{\min}(\bSigma_u)>c_1, \|\bSigma_u\|_1<c_2$ and
 $\min\limits_{i\le p, j\le p} \var(u_{it}u_{jt})>c_1.$
\item[] (iib) There are  $b_1$ and $b_2>0$ such that for any $a>0, i\le p$ and $j\le K$,
    $$P(|u_{it}|>a)\le \exp(-(a/b_1)^{2}), \ \ \ \  P(|f_{jt}|>a)\le \exp(-(a/b_2)^{2}).$$
\item[] (iic) There exists an $M>0$ such that for all $i\in J$ and  $t=1,2$ all of the quantities $\|\blb_i\|_{\max}$, $E[s_0(p)^{-1/2}\{\blu_{1J}^T\blu_{tJ}-E(\blu_{1J}^T\blu_{tJ})\}]^4$, and $E\|s_0(p)^{-1/2}\sum\limits_{i\in J} \blb_iu_{i1}\|^4$ are smaller than $M$.
\end{description}

%

\end{description}

 Recall that $\bB_{J^c\bullet}=0$ in our setting. It is easy to see that  part (i) of Condition~(A5) is weaker than the pervasive assumption  (\ref{Pervasive}) (Assumption 1 in Fan et al (2013)), which requires that $p^{-1}\lambda_{\min}(\bB^T\bB)>c>0$. Part (ii) are parallel to Assumptions 2 and 4 in Fan et al (2013). Condition~(A5) ensures that when  $(s_0(p))^{-1}\bB_{J\bullet}^T\bB_{J\bullet}>c>0$,   but $p^{-1}\bB^T\bB\rightarrow 0$, our estimate can still be consistent.
Let  $\gamma=\frac{4}{13}$ and  $m_p=\max\limits_{1\le i\le p}\sum\limits_{1\le j\le p}|\sigma_{u,ij}|^q, $ for some $q\in [0,1]$, controlling the sparsity of $\bSigma_u$.
\bet\label{theo4}
 Suppose that $\log p=o(n^{\gamma/6}), n=o(s_0(p)^2)$ and Conditions (A1)-(A3) in Subsection 2.1 and Condition~(A5) stated above  hold. Then
\beqr
\|\hat\bSigma_{u}^\tau-\bSigma_u\|&=&O_p(w_n^{1-q}m_p),\nonumber\\
\|\hat\bSigma^\tau-\bSigma\|_{\bSigma}&=&O_p\left(w_n^{1-q}m_p+p^{-1/2}s_0(p)\omega_n^2  \right),\nonumber
\eeqr
where $\|A\|_\bSigma=p^{-1/2}\|\bSigma^{-1/2}A\bSigma^{-1/2}\|_F$  defined in Fan et al (2013).
 \eet

\subsection{A comparison between PVD-based POET  and POET}

  Let $\hat\bSigma_{P}$ denote the POET estimate of $\hat\bSigma$.  Theorem 3 of Fan, et al (2013) provides that
\beq\label{eq5.3.1}
\|\hat\bSigma_P-\bSigma\|_{\bSigma}=O_p\left(\tilde w_n^{1-q}m_p+n^{-1}p^{1/2}\log p \right),
\eeq
 where $\tilde{\omega}_n=\sqrt{(\log p)/n}+\sqrt{1/p}$. First, as $0<d_1<s_0(p)/p\le d_2\le 1$ for some constant $d_1$ and $d_2$,   $\omega_n^2s_{0}(p)p^{-1/2}=O(p^{1/2}n^{-1}\log p)$ by Theorem~\ref{theo4} and   PVD-based POET  obtains exactly the same convergence rate for $\|\hat\bSigma^\tau-\bSigma\|_{\bSigma}$ as POET.   
Second,  when $s_0(p)/p\rightarrow 0$,   the signals of common factor are weak.   PVD-based POET can have better convergence rate than POET and latter may not be consistent. It is clear from (\ref{eq5.3.1}) that,
 as $p$ is large such as  $p^{1/2}n^{-1}\rightarrow \infty$,  the relative error $\|\hat\bSigma_P-\bSigma\|_{\bSigma}$ will  not converge to zero, regardless of the rate of $m_p$. This inevitably requires a strong restriction on the rate of $p$. However, for PVD-based POET method,  as long as  $s_{0}(p)=o(p^{1/2}n/\log p)$ and $n=o(s_0(p)^2)$ (the assumption required by  Theorem~\ref{theo4}), we have  $\omega_n^2s_{0}(p)p^{-1/2}=o(1)$. In this case,  the relative error $\|\hat\bSigma^{\tau}-\bSigma\|_{\bSigma}=O_p(\omega_n^{1-q}m_p)$ depends on the sparsity of $\bSigma_u$. The consistency can hold when $m_p$ is small. For example, if $m_p=o( n^{(1-q)/4})$, then by the assumption that $n=o(s_0(p)^2)$,  $\log p=o(n^{1/5})$ and the definition of $\omega_n$, it is easy to verify that $\omega_n^{1-q}m_p\rightarrow 0$. The simulation results in Section 6 confirm the conclusions here.

\section{Simulations and real data analysis}

Let $\bSigma=(\sigma_{ij})$ and $\bX_1,\cdots,\bX_n$ are i.i.d. observations from  $N_p(0, \bSigma)$. For simplicity, we take   $J=\{1,\cdots, p_1\}$ and $J^c=\{p_1+1,\cdots, p\}$.

\subsection{Pivotal variable detection}
 In this simualtion,  the sample size is $n=100$ and the dimension is $p=1000$. The experiments are repeated $T$ times  to get $\hat J_t, t=1,\cdots, T$.
 Let Mean and SD respectively stand for  the mean and standard deviation of the cardinality $|\hat J_t|$ of the set $\hat J_t$ with  $t=1,\cdots, T$;  let EQ denote the frequency of  $\hat J_t$ being exactly equal to $J$;  FP and FN respectively denote the  false positive rate and false negative rate:
$$
  EQ=\frac{1}{T}\sum\limits_{ t=1}^T \mathbf{1}_{\{\hat J_t=J\}},\ \
  FP=\frac{1}{(p-p_1)T}\sum\limits_{t=1}^T |\hat J_t\setminus J|,\ \ \
  FN=\frac{1}{p_1 T}\sum\limits_{ t=1}^T  |J\setminus \hat J_t|.
 $$
where $|\hat J\setminus J|$ denote the cardinality of the set $\hat J\setminus J$ and $|J\setminus \hat J|$ is defined similarly. In this simulation, $T=100$.
  We  consider the following two examples.

\emph{Model 1.}  Let $\bSigma=\Sigma_0^2$,  where   $\Sigma_0=(\sigma_{0,ij})$  with
$$\sigma_{0,ij}=\left\{
\begin{array}{ll}
  \rho^{2\min\{i,j\}/p_1},                 &  (i,j)\in J\times J,  \\
  \rho^{\min\{i,j\}/p_1}0.1^{2\max\{i,j\}/p},    &    (i,j)\in J^c\times J \ \mbox{or}\ (i,j)\in J\times J^c, \\
   \rho I(i\in \tilde J, j\in \tilde J),&   (i,j)\in J^c\times J^c.
\end{array}
\right.$$
  where $\tilde J\subset\{p_1+1,\cdots, p\}$ are selected at random and $|\tilde J|=30$.
  Here $\Sigma_0$ may not be positive definite, but $\bSigma$ is  positive semidefinite.

\emph{Model 2.}  Let $\bSigma=\bB\bB^T+\bSigma_u$, where $\bB=(\tilde B^T,0)^T\in R^{p\times 4}$ and  $\tilde B=(\tilde b_{ij})\in R^{p_1\times 4}$ with $\tilde b_{ij}$ independent from $N(1+\rho,0.5)$. $\bSigma_u=(\sigma_{u,ij})$ where $\sigma_{u,ij}=\rho^{|i-j|/9}I(|i-j|<9)$.

For these two models, it is easy to see that for the true covariance matrix $\bSigma$, values of $\sigma_{ij}$ in the first  $p_1$ rows and columns can be distinguished clearly from the other rows and columns. The first $p_1$ rows and columns with large  values of  $r_i$ are much denser than the others. Therefore the number of pivotal variable is $p_1$. We take different values of $\rho$ and  report the simulation results in Table~1.
  The results in this table suggest that, as $\rho$ increases from 0.1 to 0.9, the signals become stronger and  PVD can then more effectively identify the dense rows and columns in the matrix.

\begin{center}
Table~1 about here
\end{center}

\subsection{Estimation for the factor model}

Let
$\bX_i\sim N_p(0,\bSigma)$ where  $\bSigma=\bB\bB^T+\bSigma_u$. Take $J=\{1,\cdots, p_1\}$ and $\bB=(B_1^T, 0)^T\in R^{p\times 2}$, where $B_1=(b_1,\cdots, b_{p_1})^T\in R^{p_1\times 2}$ and $b_i\in R^{2}, i=1,\cdots, p_1,$ are generated which are  independent and   uniformly distributed   on the unit circle. Let $\bSigma_u=(\sigma_{u,ij})$, where $\sigma_{u,ij}=r\cdot 0.3^{|i-j|}I(|i-j|>5)$, for $1\le i,j\le p_1$; $\sigma_{u,ij}= 0.3^{|i-j|}I(|i-j|>5)$, for $p_1+1\le i,j\le p$ and $\sigma_{u,ij}=0$ otherwise. Here we use $r$ to control the significance  of $B_1B_1^T$ relative to the block matrix $\bSigma_{u,JJ}$.
Larger $r$ means the clearer differences between  low rank matrix and sparse one and consequently easier  to separate them.
Let $\hat\bSigma^{\tau}$ and $\hat\bSigma_u^{\tau}$ respectively denote  the estimates of $\bSigma$ and $\bSigma_u$. To simplify the comparison,   we report  the relative error $\mathrm{RE}=p^{-1/2}\|\bSigma^{-1/2}\hat\bSigma^\tau\bSigma^{-1/2}-I_p\|_F$ (see, Fan et al, 2013)  and $\mathrm{EU}=\|\hat\bSigma_u^\tau-\bSigma_u\|$ for all the competitors.
Set $r=0.1, 0.5, 1$  respectively.

\subsubsection{Comparison between LOREC  and PVD-based LOREC}

  Consider several  configurations of $p$ and $p_1$. The performance of PVD is similar to that with Model~2 in the previous subsection and thus the results are not reported here for conciseness. We repeat replica 100 times to compute the RE and EU.  The simulation results of LOREC and PVD-based LOREC  are presented in Table~2. Besides, we also report the average CPU time  in seconds for one experiment in the replications, denoted by TM, in a working station with Intel(R) Xeon(R) CPU E5 2603 1.80GHz.

\begin{center}
Table~2 about here
\end{center}

From Table~2, we have several observations.  First, the simulation results  obviously show that, compared with PVD-based LOREC, the computation of LOREC is  very intensive even when $p_1$ ($p_1=20$ say) is much smaller than $p$.  This is because LOREC is actually a general method and thus has no advantage for sparse factor model. This is also the reason that LOREC  cannot handle large $p$ cases in practice and theory. In this case, the computational efficiency of PVD-based LOREC is very significant because the PVD step can make the working dimension much smaller than the original $p$ such that PVD-based LOREC works efficiently in computation. For example, when $p=300, p_1=20$, PVD-based LOREC uses less than 9 seconds per experiment on average whereas   LOREC uses more than  2700 seconds that is 300 times more  than that of PVD-based LOREC.
When $p_1$ is large, such as $p_1=90$ or 120, PVD-based LOREC uses much more time, in other words, PVD can reduce the original dimension $p$ less. But even though PVD is still helpful.  This means that the computational time of the PVD step is negligible compared with the LOREC step. Second,   PVD-based LOREC performs much better than LOREC, especially when  $r$ is small such as $0.1$. We note that in this case, the signal of sparse matrix $\bSigma_{u,JJ}$ is weaker and it is  difficult to separate it from the low rank matrix. Thus, LOREC cannot work well. Moreover, given  $p_1$, the performance of  PVD-based LOREC are  stable for  different $p$ whereas, as $p$ increases,  LOREC   becomes worse as expected. This further suggests the usefulness of the PVD step.  Finally, under the large $p_1$ cases such as $90$ or $120$, LOREC  can work better than that  under the small $p_1$ cases such as $p_1=20$. This is because of the  increase of the signal of low rank  matrix.

\subsubsection{Comparison between POET and PVD-based POET}
As POET can handle large $p$ cases, therefore, in this comparison, we consider larger $p$ than those in the previous subsection. Furthermore, the values of  $p=300+100 \times i$ for $i=0, \cdots, 7$ are taken to check the dimensionality influence on the estimation efficiency. We then do not report the detail of the average  CPU time here. Also, by theory, POET works when $p_1$ is not too small. Thus, to compare with POET and PVD-based POET, we set  $p_1=120$, $n=150$. The performance of the PVD step is similar to that under Model~2 in the previous subsection and again the results are not reported here.  First, we find that PVD-based POET uses about 70\% of the workload that POET uses. In other words, POET is much more computational efficient than LOREC when we compare the results under the cases with $p=200$ and $300$. Figure~1 presents the mean of relative error $\mathrm{RE}$ (in plots (a)--(c))  and $\mathrm{EU}$ (in plots (d)--(f)) over 100 replicas.    In Step (3) of  PVD-based POET in Section~5,   the thresholding values $0.5\hat\theta_{ij}^{1/2}[ (\log p/n)^{1/2}+s_0(p)^{-1/2}]$  are used (see Fan et al (2013)).

\begin{center}
Figure~1 about here
\end{center}


 The results indicate that when $p$ is relatively small $p\le 500$, PVD-based POET performs similarly as  POET for all $r$. However, when $p$ is large ($p\ge 500$),  PVD-based POET is clearly the winner.  When $p$ gets larger, the impact  from the common factors significantly decreases. The space spanned by the larger eigenvectors of $\hat\bSigma$ does not converge to that spanned by the columns of $\bB$. Consequently, for nonsparse factor model,   span$(\bB)$ cannot be  estimated well by the space spanned by the eigenvectors of the sample covariance matrix $\hat\bSigma$ obtained by  POET. As a result, $\blu_i$ cannot be consistently estimated. This  causes the poor performance of the POET-based estimates of $\bSigma_u$ and $\bSigma$. From Figure~1, we can see that the POET estimates have much larger RE and EU  than the PVD-based POET estimates under the large $p$ cases.

Further,  it is observed that as $r$ decreases, POET causes larger RE.  The main reason is that for small $r$,  $\bSigma$ is close to singular, that is, the condition number of $\bSigma$ is large. Since the POET-based estimate $\hat\bSigma$ is  inconsistent to $\bSigma$.  the relative error(RE) that  involves $\bSigma^{-1}$ is amplified in  small $r$ cases. However, for the sparse factor model in the simulations, we see that the relative error of the PVD-based POET estimate is stable to both $r$ and $p$. Therefore, PVD-based POET performs well in the case of $\bSigma$ being close to singular. On the other hand, we see that for all $r$, the average EU values  of POET retain much larger than those of PVD-based POET when $p$ is large.

Finally, in the case of  $p_1=120, p=300$, we can compare the simulation results of  POET and  PVD-based POET here with those of LOREC and  PVD-based LORE in  Table 2.
 In terms of EU, it is easy to see that   LOREC and PVD-based LOREC are  much worse than   POET  and PVD-based POET accordingly. Note that LOREC uses  $L_1$ penalty in estimating  $\bSigma_u$, while POET uses an adaptive estimate of $\bSigma_u$ (Cai and Liu, 2011; Fan et al, 2013). This could be a main reason. Further,  LOREC causes larger RE than the other three competitors when $r=0.1$. When $r=0.5$,  all the methods are similar, and for $r=1$ LOREC and PVD-based LOREC are slightly better than POET and PVD-based POET accordingly and PVD-based LOREC is the best.

\subsection{Real data analysis}

The purpose of this analysis is to examine how PVD can efficiently  help on a sparse factor modelling and estimation. We consider a Gliobastoma microarray gene expression data set from the Cancer Genome Atlas Project (https://tcga-data.nci.nih.gov/tcga/). The level 3 summarized data were downloaded,
and then batch effects were corrected with combat (Johnson
et al., 2007). This data set was used in the joint analysis of micro-RNA and RNA data in Chen et al. (2013).
It contains 12042 genes and 484 observations. Our purpose of using this data set is to examine whether PVD can effectively detect pivotal variables  such that the LOREC- and POET-based estimate can work better. To this end, we first select genes with the standard deviations(SD) between 1 and 1.5. There are  4544 genes retained.
To check whether PVD can perform stable for this data set, we  select 250 observations at random each time and run PVD to select the pivotal genes. The process is repeated $T=50$ times. The average number of pivotal variables and the corresponding standard deviation are 10.16 and 7.56, respectively.
 The numbers of the genes selected  in 50 times are presented in Figure~2. It can be inferred that the number of  selected pivotal genes is relatively  stable.

\begin{center}
Figure~2 about here
\end{center}


 Therefore, we start to perform PVD. First, we further consider genes with the first 200($p=200$) largest  standard deviations from those genes whose SD is smaller than 1.5. The four methods: LOREC,  PVD-based LOREC, POET and PVD-based POET are performed. When considering factor modelling,
 LOREC finds 10 common factors and the corresponding covariance matrix is similar to but slightly  sparser than the ordinary sample covariance. When PVD  is used, 47 pivotal genes are detected from these 200 genes, and then PVD-based LOREC finds 3 common factors. The corresponding covariance matrix is reasonably sparser than that obtained by LOREC. For both POET and PVD-based POET, only  one common factor is considered and the corresponding covariance matrices  are sparser than LOREC and PVD-based LOREC find.   The corresponding  heatmaps of the sample covariance matrix, and the estimated covariance matrices by LOREC,  PVD-based LOREC, POET and PVD-based POET are respectively presented in
Figures 3-7.  It is clear that PVD helps on estimation and PVD-based POET can get sparser solution than all the competitors.

\begin{center}
Figures~3-7 about here
\end{center}

%
%
%
%
%
%
%
%


\section{Appendix}
This subsection contains the proofs of the  theorems 2 and 3 and   proofs  of other theorems are provided in Supplementary materia.

%

{\it Proof of Theorem  \ref{prop1}} \, \,
 Theorem 1 shows that under Condition (A2) and (A3),
$  \max\limits_{1\le i\le p}|\hat r_i-r_i|\le C_0q_n$,   with  a probability $1-O(p^{-\delta_0})$.
Therefore, as $n \to \infty$ , with a probability tending to 1, we have    $\max\limits_{j\in J^c} \hat r_i<\delta_p+C_0q_{n} $ and  $\min\limits_{j\in J} \hat r_i\ge c_p-C_0q_{n}$. As $q_{n}\rightarrow 0$,  it holds that $\delta_p/q_n\rightarrow 0$ and  $c_p/q_n\rightarrow \infty$ by Conditions (A1).  Consequently, we have $\max\limits_{j\in J^c} \hat r_i/\min\limits_{j\in J} \hat r_i=o_p(1)$.  This completes the proof. $\blacksquare$

{\it Proof of Theorem  \ref{theo3}} \, \,
  Under Conditions~(A1)-(A3), Theorem  1 shows that, with probability tending to 1,
\beq\label{gap}
\max_{1\le i\le p}|\hat r_i-r_i|<C_0q_n.
\eeq

Suppose that $J=\{1,\cdots, s_0(p)\}=\cup_{m=1}^M J_m$, such that for  each $J_m$,  $r_i$ with $i\in J_m$ takes  the same value.  Noting that $r_i$'s are arranged  in the descending  order, we assume $J_m=\{j_{m-1}+1,\cdots, j_m\}$ where  $0=j_0<j_1<\cdots< j_M=s_0(p)$.  For $1\le i\le p$, let $(i)$ denote the index $i_0$ such that $\hat r_{i_0}=\hat r_{(i)}$. Then define the sets $\hat J_m=\{(j_{m-1}+1),\cdots, (j_m)\}$, $m=1,\cdots, M$.
By Condition (A1), $\max\limits_{i\in J} r_i/\min\limits_{i\in J} r_i =O(1) $ and $r_i=O(c_p), i\in J$ uniformly. Then   we have for some  $0<m_1< m_2< \infty$, such  that $  m_1<r_i/c_p<m_2$  or $m_1c_p<r_i<m_2c_p$ for any $1\le i\le s_0(p)$.  Then together with  (\ref{gap}), we have
 $P(m_1c_p-C_0q_n <\hat r_{i}\le m_2c_p+C_0q_n,  1\le i\le s_0(p))\rightarrow 1.$
By Condition~(A1),  we have  $\max\limits_{i\in J^c} r_i \le m_3 \delta_p$ for some $0<m_3<\infty$.
Then (\ref{gap}) yields that
$P(\max_{i>s_0(p)}\hat r_{i}\le  m_3\delta_p+C_0q_n)\rightarrow 1.$
 Let  $A_n=\{\cup_{m=1}^M\hat J_m=\cup_{m=1}^M J_m\}$. Combing the
 two formulas above with the fact that $\max(\delta_p,q_n)=o(c_p)$,   we have $P(A_n)\rightarrow 1$, that is,  the index set  $\{(i), 1\le i\le s_0(p)\}$ are consistent estimate of $J$.

Next  we estimate  $s_0(p)$.
By the   definitions of $l_n$ in Theorem 3, $c_p$ and $q_n$ in Condition (A1) and the fact that $\hat r_{(i)}>0$, as $n\rightarrow \infty$,
it follows that  with a probability tending to 1
$$ \min\limits_{1\le i< s_0(p)}    R_i= \min\limits_{1\le i< s_0(p)} \frac{\hat r_{(i+1)}+l_n}{\hat r_{(i)}+l_n}\ge \frac{m_1c_p-C_0q_n +l_n}{m_2c_p+C_0q_n+l_n}\rightarrow    m_1/m_2>0,$$
$$    R_{s_0(p)}=\frac{\hat r_{(s_0(p)+1)}+l_n}{\hat r_{(s_0(p))}+l_n}<\frac{m_3 \delta_p+C_0q_n +l_n}{m_2c_p+C_0q_n+l_n}\rightarrow 0,$$
$$   \min\limits_{ i> s_0(p)}  R_{i}=  \min\limits_{ i> s_0(p)}\frac{\hat r_{(i+1)}+l_n}{\hat r_{(i)}+l_n}\ge \frac{l_n}{m_3\delta_p+C_0q_n+l_n}\rightarrow 1,$$
where we have used the fact $\max(\delta_p, q_n)=o(l_n)$ and $l_n=o(c_p)$.
Combining the above results with the definition of $\hat s_0(p)$ in (\ref{eq2.1}), we have  $P(\hat s_0(p)=s_0(p))\rightarrow 1.$
 Further, recall that $P(A_n)\rightarrow 1$ and that   $\hat J=\{i: \hat r_i\ge \hat r_{(\hat s_0(p))}\}=\{(i):  1\le i\le \hat s_0(p)\}$. Thus,  we have $P(\hat J=J)\rightarrow 1$.
  This completes the proof. $\blacksquare$

\renewcommand{\baselinestretch}{1}
\normalsize
\section*{References}
\begin{description}
\newcommand{\enquote}[1]{``#1''}
\expandafter\ifx\csname natexlab\endcsname\relax\def\natexlab#1{#1}\fi

\bibitem[{Agarwal et al.(2012)}]{Agarwal:2012}  Agarwal, A., Negahban, S., and  Wainwright, M. J. (2012). Noisy matrix decomposition via convex relaxation: Optimal rates in high dimensions. {\it  Annals of Statistics}, {\bf 40}, 1171-1197.

\bibitem[{Bai and Yin(1993)}]{Bai:1993}  Bai, Z. D., and  Yin, Y. Q. (1993). Limit of the smallest eigenvalue of a large dimensional sample covariance matrix. {\em  Annals of Probability}, {\bf 21},  1275--1294.

\bibitem[{Bai and Ng(2002)}]{Bai:2002} Bai, J., and Ng, S. (2002). Determining the number of factors in approximate factor models. {\it Econometrica}, {\bf 70}, 191-221.

\bibitem[{Bai(2003)}]{Bai:2003}  Bai, J. (2003). Inferential theory for factor models of large dimensions. {\it Econometrica}, {\bf 71}, 135-171.


\bibitem[{Bai and Li(2012)}]{Bai:2012} Bai, J., and Li, K. (2012). Statistical analysis of factor models of high dimension. {\it The Annals of Statistics}, {\bf 40}, 436-465.


\bibitem[{Bickel and Levina (2008)}]{Bickel:2008}   Bickel, P. and   Levina, E. (2008). Covariance regularization by thresholding.  {\em   Annals of Statistics}  {\bf 36}, 2577--2604.


\bibitem[{Cai and Liu(2011)}]{Cai:zhou:2012} Cai, T. T. and Liu, W. D. (2011). Adaptive thresholding for sparse covariance matrix estimation.   {\em   Journal of the American Statistical Association},  {\bf 106},   672--684.


\bibitem[{Cai and Zhou (2012)}]{Cai:zhou:2012} Cai, T. T. and Zhou, H. H. (2012). Optimal rates of convergence for sparse covariance matrix estimation.   {\em  The  Annals of Statistics},  {\bf 40}, 2389--2420.


\bibitem[Chandrasekaran et al(2012)]{Chandrasekaran:2012}  Chandrasekaran, V., Parrilo, P. A., nad Willsky, A. S. (2012). Latent Variable Graphical Model Selection via Convex Optimization. {\it The Annals of Statistics}, {\bf 40}, 1935-1967.


\bibitem[{Chen et al. (2013)}]{Chen:etal:2013} Chen, X., Slack, F. J. and Zhao, H.  (2013). Joint analysis of expression profiles from multiple cancers improves the identification of microRNA-gene interactions. {\it Bioinformatics}, {\bf 29}, 2137--2145.

\bibitem[{Fan et al (2008)}]{Fan:2008}  Fan, J., Fan, Y., and Lv, J. (2008). High dimensional covariance matrix estimation using a factor model. {\it Journal of Econometrics}, {\bf 147}, 186--197.

\bibitem[{Fan et al. (2011)}]{Fan:etal:2011}  Fan, J., Liao, Y., and Mincheva, M.  (2011).  High-dimensional covariance matrix estimation in approximate factor models.  {\em  The  Annals of Statistics},  {\bf 39},  3320--3356.

\bibitem[{Fan et al (2013)}]{Fan:etal:2013}  Fan, J., Liao, Y., and Mincheva, M. (2013).  Large covariance estimation by thresholding
principal orthogonal complements {\em Journal of  Royal Statistic  Socociation, Series B}, {\bf 75},  1--44.

\bibitem[{Johnson et al. (2007)}]{Johnson:etal:2007} Johnson, W. E., Li, C., and Rabinovic, A. (2007). Adjusting batch effects in microarray expression data using empirical Bayes methods. {\it Biostatistics}, {\bf 8}, 118--127.

\bibitem[{Johnstone (2001)}]{Johnstone:2001}  Johnstone, I. M. (2001). On the distribution of the largest eigenvalue in principal components analysis. {\it The Annals of Statistics}, {\bf 29}, 295--327.

\bibitem[{Luo(2011)}]{Luo:2011} Luo, X. (2011). High Dimensional Low Rank and Sparse Covariance Matrix Estimation via Convex Minimization. {\it   arXiv}  1111.1133.


\bibitem[{Ravikumar et al (2011)}]{Ravikumar:2011} Ravikumar, P., Wainwright, M. J., Raskutti, G., and Yu, B. (2011). High-dimensional covariance estimation by minimizing $\ell_1$-penalized log-determinant divergence.  {\it Electronic Journal of Statistics}, {\bf 5}, 935--980.

\bibitem[{Rothman et al (2009)}]{Rothman:2009} Rothman, A. J., Levina, E., and Zhu, J. (2009). Generalized thresholding of large covariance matrices. {\em Journal of the American Statistical Association},  {\bf 104}, 177--186.

\bibitem[{Serfling (1980)}]{Serfling:1980} Serfling, R. J. (1980). Approximation theorems of mathematical statistics. {\it Wiley series in probability and mathematical statistics.}

\bibitem[{Stock and Watson(1998)}]{Stock:1998}  Stock, J. H., and Watson, M. W. (1998). Diffusion indexes {\em Working Paper 6702}. National Bureau of Economic Research, Cambridge.

\bibitem[{Stock and Watson(2002)}]{Stock:2002} Stock, J. H., and Watson, M. W. (2002). Forecasting using principal components from a large number of predictors. {\it Journal of the American statistical association}, {\bf 97}, 1167-1179.


\bibitem[{Vershynin(2011)}]{Vershynin:2011} Vershynin, R. (2011). {Introduction to the non-asymptotic analysis of random matrices}. \text{arXiv:1011.3027v5}.


\bibitem[{Xia et al(2013)}]{Xia:2013} Xia, Q., Xu, W. L., and Zhu, L. X. (2014). Consistently determining the number of factors in multivariate
volatility modelling. {\it Statistica Sinica}, accepted.

\end{description}

\newpage

\begin{table}[h] \caption{Simulation results for $p=1000$}
  \centering
\begin{tabular}{lllrrrrr}
  \hline
    model   &      $p_1$  &$\rho$& Mean  & SD  &FP    &   FN  &EQ \\\hline
            &             &0.1   &42.00 &15.08 & 0.00 &  0.16 &0.40  \\
            &             &0.3   &50.07 & 0.30 & 0.00 &  0.01 &0.94   \\
            &    50       &0.5   &50.01 & 0.10 & 0.00 &  0.00 &0.99   \\
            &             &0.7   &50.01 & 0.10 & 0.00 &  0.00 &0.99   \\
            &             &0.9   &50.01 & 0.10 & 0.00 &  0.00 &0.99   \\ \cline{2-8}

            &             &0.1   & 65.42 &44.92 & 0.00 & 0.35 &0.19   \\
            &             &0.3   & 99.26 &10.00 & 0.00 & 0.01 &0.78   \\
     (1)      &   100       &0.5   &100.02 & 0.31 & 0.00 & 0.00 &0.93   \\
            &             &0.7   &100.10 & 0.48 & 0.00 & 0.00 &0.94   \\
            &             &0.9   &100.03 & 0.26 & 0.00 & 0.00 &0.95   \\ \cline{2-8}

            &             &0.1    & 117.51 &109.73 & 0.01 & 0.45&0.02  \\
            &             &0.3    & 172.49 & 69.40 & 0.00 & 0.13&0.53  \\
            &    200      &0.5    & 198.38& 10.95 & 0.00 & 0.01&0.78  \\
            &             &0.7    & 199.56 & 1.05 & 0.00 & 0.01&0.94  \\
            &             &0.9    & 200.10 & 0.30 & 0.00 & 0.00&0.98 \\ \hline

          &             &0.1    &  40.69 &   19.31 & 0.00 &  0.18 &  0.81       \\
          &             &0.3    &  48.35 &    8.34 & 0.00 &  0.03 &  0.90        \\
          &    50       &0.5    &  50.00 &    0.00 & 0.00 &  0.00 &  1.00        \\
          &             &0.7    &  50.00 &    0.00 & 0.00 &  0.00 &  1.00        \\
          &             &0.9    &  50.00 &    0.00 & 0.00 &  0.00 &  1.00        \\ \cline{2-8}

          &             &0.1    &  82.75 &   37.18 &  0.00 &  0.17 &  0.41        \\
          &             &0.3    &  98.95 &    9.79 &  0.00 &  0.01 &  0.94        \\
   (2)      &   100       &0.5    & 100.00 &    0.00 &  0.00 &  0.00 &  1.00        \\
          &             &0.7    & 100.00 &    0.00 &  0.00 &  0.00 &  1.00        \\
          &             &0.9    & 100.00 &    0.00 &  0.00 &  0.00 &  1.00        \\ \cline{2-8}

          &             &0.1    & 122.12 &  96.28  &  0.00 &  0.38 &  0.24      \\
          &             &0.3    & 199.38 &   0.48  &  0.00 &  0.00 &  0.38      \\
          &    200      &0.5    & 200.00 &   0.00  &  0.00 &  0.00 &  1.00      \\
          &             &0.7    & 200.00 &   0.00  &  0.00 &  0.00 &  1.00      \\
          &             &0.9    & 200.00 &   0.00  &  0.00 &  0.00 &  1.00      \\ \hline
\end{tabular}

\end{table}

\newpage

\begin{table}[h] \caption{LOREC and PVD-based LOREC with the sample size $n=150$}
  \centering
\begin{tabular}{lllrrr|cccc}\hline
   &  &    &      \multicolumn{3}{c|}{LOREC}   &  \multicolumn{3}{c}{PVD-based LOREC}\\   \cline{4-9}
\multirow{2}{*}{$p$}  &\multirow{2}{*}{$p_1$}&  & \multicolumn{3}{c|}{$r$} & \multicolumn{3}{c}{$r$} \\
 &       &      &            0.1 &  0.5 &  1&  0.1 &  0.5&  1 \\\hline

      &20 & EU  & 25.739 & 21.885 & 18.744 & 15.651 & 10.879 &8.653  \\
      &   & RE  & 1.141  &0.570   & 0.482  & 0.847 & 0.506  &0.469\\
      &   & TM  & 357.175&339.088  &342.725  &8.092  &8.070  &7.899\\

100&90 & EU  & 18.632      & 13.669 & 11.908 & 6.452 & 4.335 & 1.533   \\
   &   & RE  &1.053    & 0.502   &0.450    & 0.801    &0.479    & 0.448   \\
   &   & TM  & 317.930 & 347.613 &365.470  & 234.173  & 248.974 &267.590\\
   &   &     &  &  &  &  &  &\\

   &20  & EU  & 32.825 & 25.356 &21.172  & 15.935 & 10.298 &7.772\\
   &    & RE  & 1.463  &0.607   & 0.534  & 0.772  & 0.516  &0.490\\
   &   & TM  & 1899.971& 1759.109 &2273.945  &8.717  &7.919  &7.853\\

200&120  & EU & 30.485  & 10.238 &5.651  &9.189  &4.033  &3.534  \\
   &    & RE & 1.242 & 0.571 & 0.514 &0.811  &0.429  &0.411  \\
   &   & TM  &1770.551 & 1766.879 &1896.734  &443.369  &446.668  &486.941\\
   &   &     &  &  &  &  &  &\\

     &20  & EU & 35.565 & 25.956 &22.166  &14.773  & 9.121 & 8.057 \\
     &    & RE &1.591  &0.627  &0.572  & 0.709 & 0.522 &0.483  \\
     &   & TM  &2750.761   & 2744.443 & 2894.143 &8.842  & 8.742  & 8.537\\

  300&120  & EU & 32.152  & 11.040 & 6.036 &8.709  &3.847  &3.605  \\
     &    & RE & 1.368 &0.597  &0.551  &0.673  &0.463  & 0.454 \\
     &   & TM  &5238.465  &5430.726  &6050.430  & 457.111 &511.934  &509.842\\
     \hline
\end{tabular}
\end{table}

\newpage

\begin{figure}
\begin{center}
\includegraphics[width=6in]{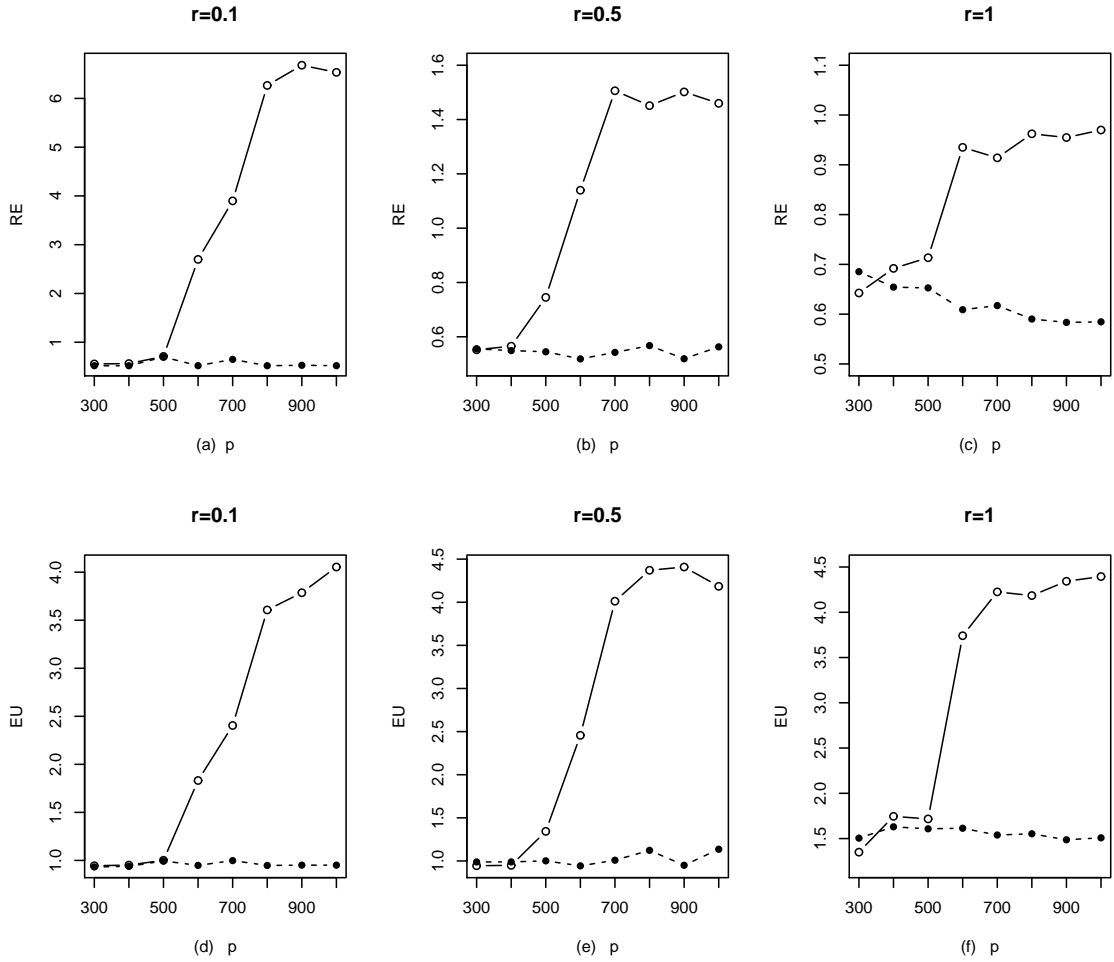}
\caption{  Plots (a)--(c) present  the  average   RE  and plot  (d)--(f) present the average EU for different values of $r$. In each plot, the solid line represents the results for POET and the dotted line   for  PVD-based POET. }\label{fig:1}
\end{center}
\end{figure}

\begin{figure}
\begin{center}
\includegraphics[width=4in]{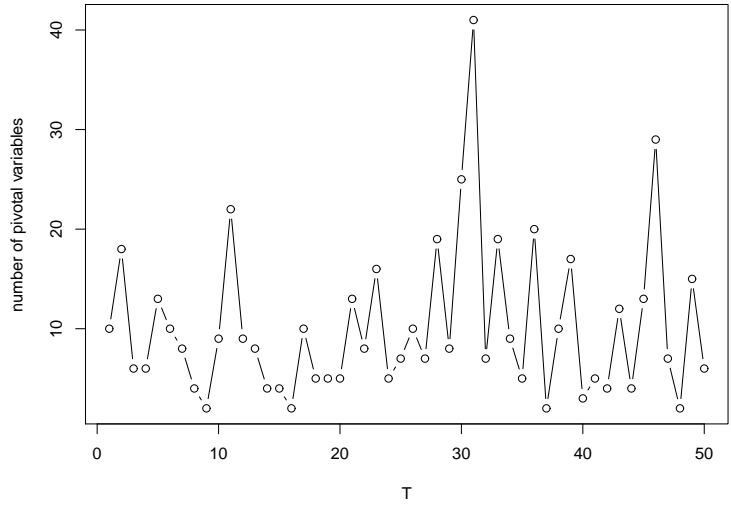}
\caption{ The number of pivotal genes selected in T=50 replicas}
\end{center}
\end{figure}

\newpage

\begin{figure}
\begin{center}
\includegraphics[width=3in]{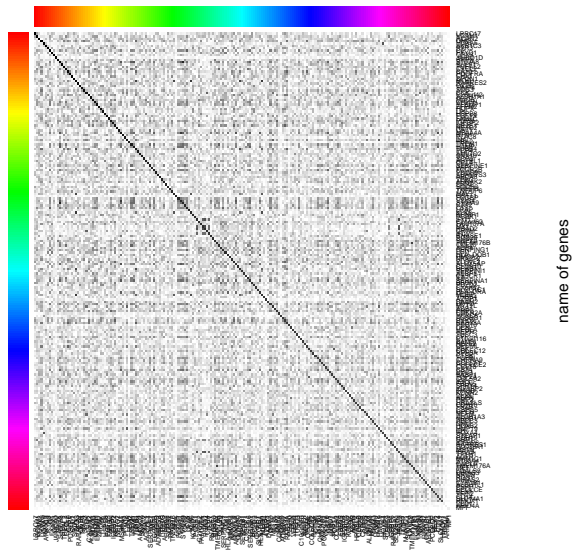}
\caption{ Sample covariance matrix}
\end{center}
\end{figure}

\begin{figure}
\begin{center}
\includegraphics[width=3.5in]{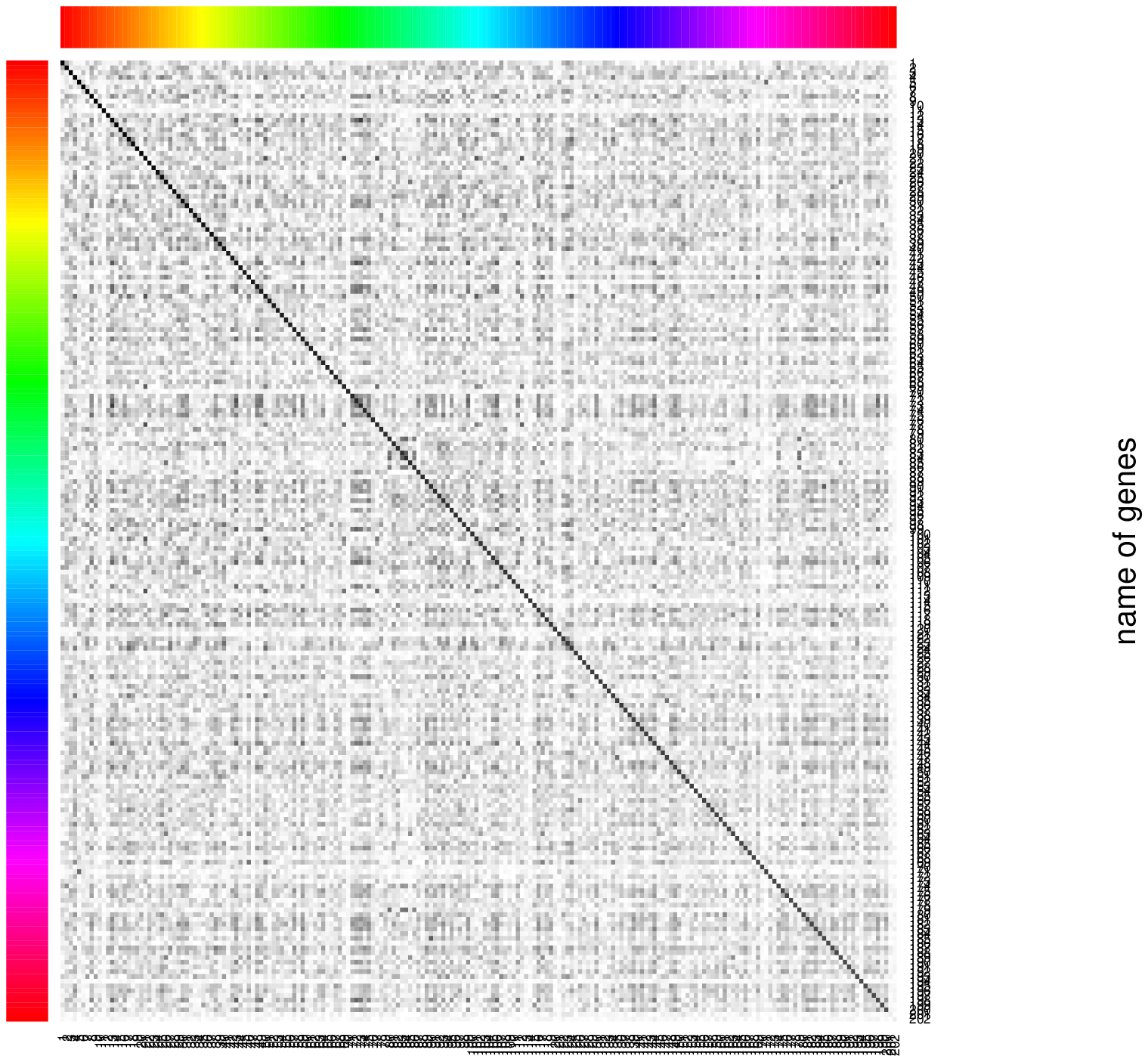}
\caption{ Estimated covariance matrix   by LOREC}
\end{center}
\end{figure}

\begin{figure}
\begin{center}
\includegraphics[width=3in]{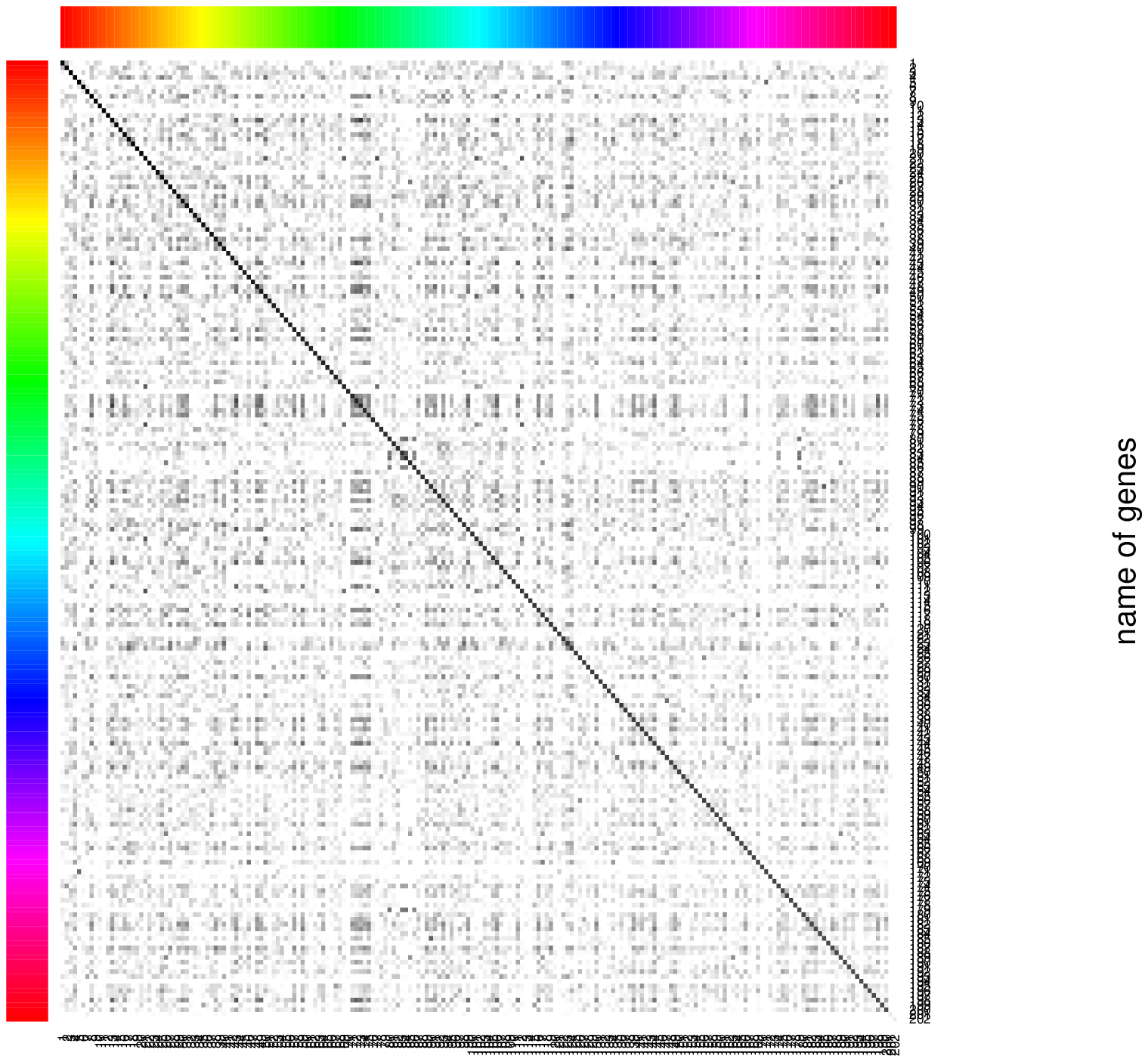}
\caption{ Estimated covariance matrix   by PVD-based LOREC}
\end{center}
\end{figure}

\begin{figure}
\begin{center}
\includegraphics[width=3in]{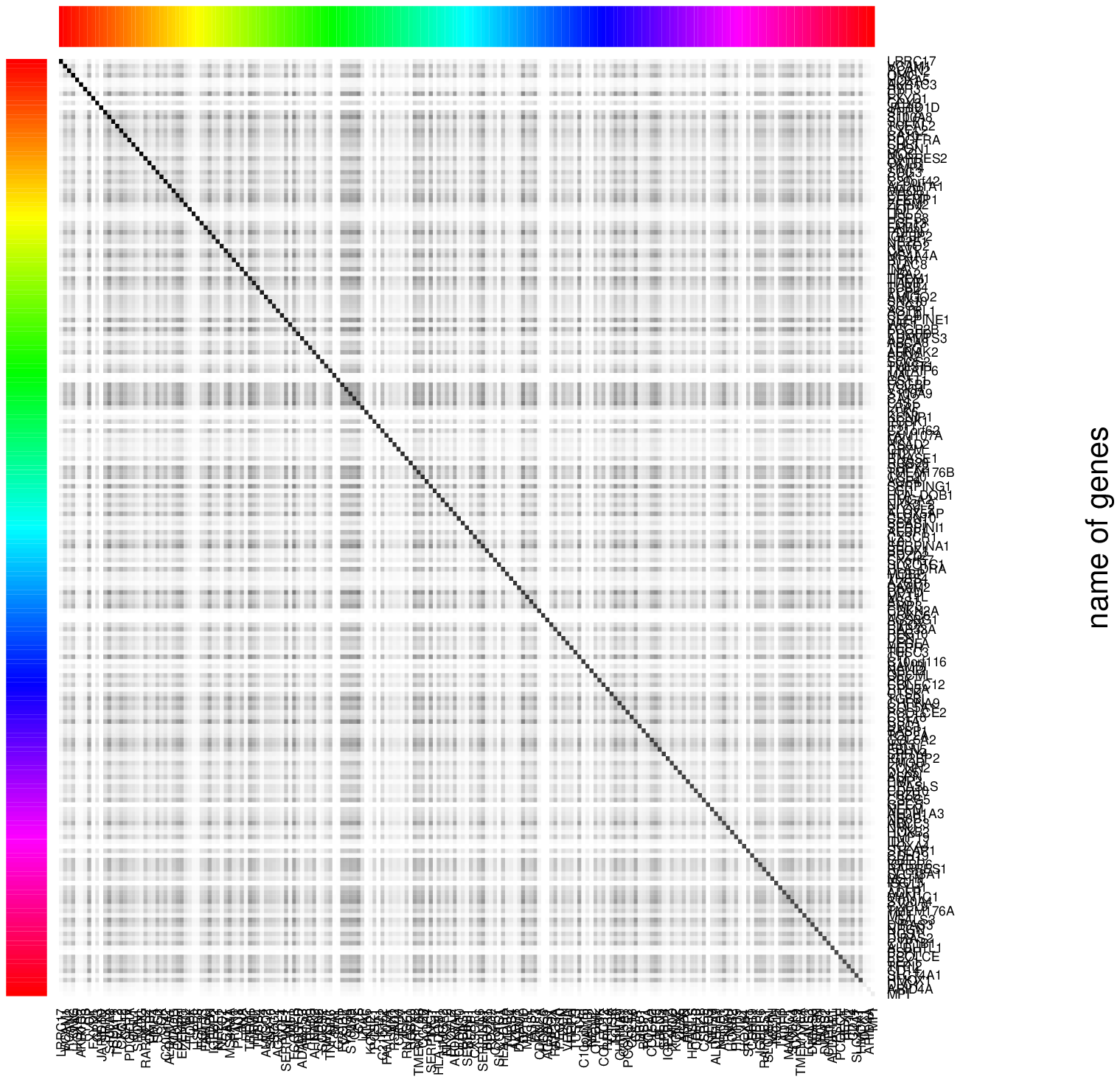}
\caption{ Estimated covariance matrix   by  POET}
\end{center}
\end{figure}

\begin{figure}
\begin{center}
\includegraphics[width=3in]{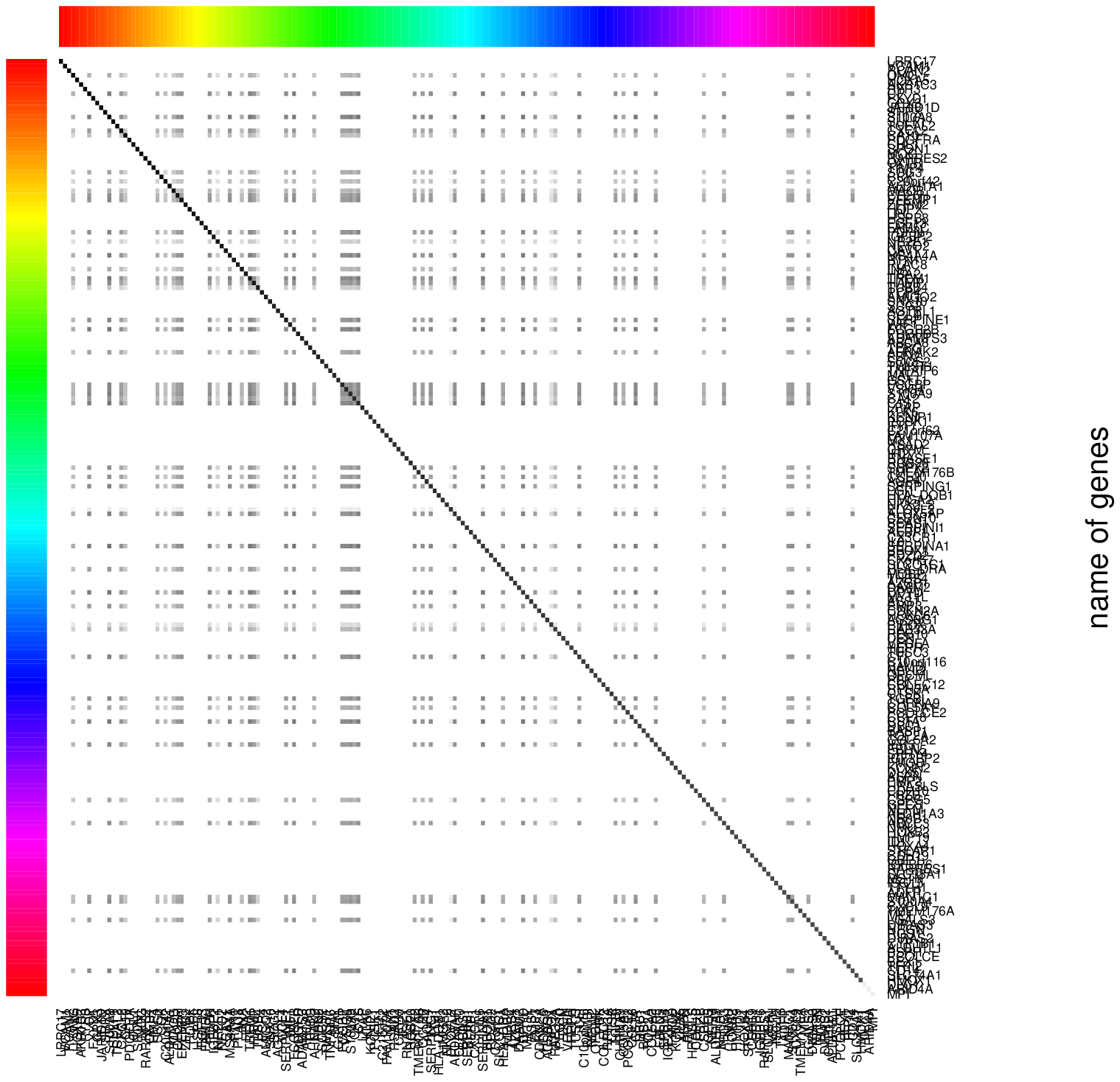}
\caption{ Estimated covariance matrix   by PVD-based POET}
\end{center}
\end{figure}

\end{document}